\documentclass[a4paper,twoside,reqno,11pt,tbtags]{article}

\usepackage{amssymb,amsmath,amsthm,mathrsfs}
\usepackage{graphicx}
\usepackage{times}

\leftmargini=\baselineskip

\newtheoremstyle{localthm}
	{7pt} 
	{7pt} 
	{\sl} 
	{} 
	{\bf} 
	{{\rm.}} 
	{.7em} 
	{} 

\theoremstyle{localthm}
\newtheorem{Theorem}{Theorem}
\newtheorem{Corollary}[Theorem]{Corollary}

\newtheorem{Lemma}[Theorem]{Lemma}

\newtheoremstyle{localrem}
	{5pt} 
	{5pt} 
	{\rm} 
	{} 
	{\bf} 
	{{\rm.}} 
	{.7em} 
	{} 

\theoremstyle{localrem}
\newtheorem*{Definition}{Definition}
\newtheorem*{Remark}{Remark}
\newtheorem*{Example}{Example}

\def\Ex{\mathop{\mathrm{I\!E}}\nolimits}
\def\Pr{\mathop{\mathrm{I\!P}}\nolimits}

\newcommand{\R}{\mathbb{R}}

\newcommand{\NN}{\mathcal{N}}
\newcommand{\FF}{\mathcal{F}}
\newcommand{\FFblc}{\FF_{\rm blc}}

\def\hat{\widehat}
\def\tilde{\widetilde}

\def\eps{\varepsilon}

\begin{document}
\addtolength{\baselineskip}{0.5\baselineskip}

\title{Bi-Log-Concave Distribution Functions$^*$}
\author{Lutz D\"umbgen$^1$, Petro Kolesnyk$^1$ and Ralf A.\ Wilke$^2$\\
	($^1$University of Bern and $^2$Copenhagen Business School)}
\date{October 2016}

\maketitle

\begin{abstract}
Nonparametric statistics for distribution functions $F$ or densities $f=F'$ under qualitative shape constraints constitutes an interesting alternative to classical parametric or entirely nonparametric approaches. We contribute to this area by considering a new shape constraint: $F$ is said to be bi-log-concave, if both $\log F$ and $\log(1 - F)$ are concave.  Many commonly considered distributions are compatible with this constraint. For instance, any c.d.f.\ $F$ with log-concave density $f = F'$ is bi-log-concave. But in contrast to log-concavity of $f$, bi-log-concavity of $F$ allows for multimodal densities. We provide various characterisations. It is shown that combining any nonparametric confidence band for $F$ with the new shape constraint leads to substantial improvements, particularly in the tails. To pinpoint this, we show that these confidence bands imply non-trivial confidence bounds for arbitrary moments and the moment generating function of $F$.
\end{abstract}

\bigskip

\noindent
$^*$Work supported by Swiss National Science Foundation.

\paragraph{AMS subject classifications.}
62G15, 62G20, 62G30.

\paragraph{Key words.}
hazard, honest confidence region, moment generating function, moments, reverse hazard, shape constraint.

\section{Introduction}

In nonparametric statistics one is often interested in estimators or confidence regions for curves such as densities or regression functions. Estimation of such curves is typically an ill-posed problem and requires additional assumptions. Interesting alternatives to smoothness assumptions are qualitative constraints such as, for instance, monotonicity or concavity.

Estimation of a distribution function $F$ based on independent, identically distributed random variables $X_1, X_2, \ldots, X_n$ with c.d.f.\ $F$ is common practice and does not require restrictive assumptions. But nontrivial confidence regions for certain functionals of $F$ such as the mean do not exist without substantial additional constraints (cf.\ \nocite{Bahadur_Savage_1956}{Bahadur and Savage, 1956}).

A growing literature on density estimation under shape constraints considers the family of log-concave densities. These are probability densities $f$ on $\R^d$ such that $\log f : \R^d \to [-\infty,\infty)$ is a concave function. For more details see \nocite{Bagnoli_Bergstrom_2005}{Bagnoli and Bergstrom (2005)}, \nocite{Cule_etal_2010}{Cule et al.\ (2010)}, \nocite{Duembgen_Rufibach_2009, Duembgen_Rufibach_2011}{D\"umbgen and Rufibach (2009, 2011)}, \nocite{Walther_2009}{Walther (2009)}, \nocite{Seregin_Wellner_2010}{Seregin and Wellner (2010)}, \nocite{Duembgen_etal_2011}{D\"umbgen et al.\ (2011)} and the references cited therein. Most efforts in these papers are devoted to point estimation. \nocite{Schuhmacher_etal_2011}{Schuhmacher et al.\ (2011)} obtain a nonparametric confidence region by combining the log-concavity constraint and a standard Kolmogorov-Smirnov confidence region. But its explicit computation is difficult, and this is one motivation to search for alternative shape constraints in terms of the distribution function $F$ directly.

While many popular densities are log-concave, this constraint can be too restrictive in applications with a multimodal density. In the present paper we consider a model with a new and weaker constraint on the distribution function:

\begin{Definition}[Bi-log-concavity]
A distribution function $F$ on the real line is called \textsl{bi-log-concave} if both $\log F$ and $\log(1 - F)$ are concave functions from $\R$ to $[-\infty,0]$.
\end{Definition}

\noindent
Many distribution functions satisfy this constraint. In particular, when $F$ has a log-concave density $f = F'$, it is bi-log-concave \nocite{Bagnoli_Bergstrom_2005}{(Bagnoli and Bergstrom, 2005)}. But indeed, bi-log-concavity of $F$ is a much weaker constraint. As shown later, $F$ may have a density with an arbitrary number of modes. Thus, we consider estimation of distributions under shape constraints for a wider family of distributions.

The remainder of this paper is organized as follows: In Section~\ref{sec:Bi-log-concave} we present characterisations of bi-log-concavity and explicit bounds for $F$ and its density $f = F'$. In Section~\ref{sec:Confidence bands} we describe exact (conservative) confidence bands for $F$. They are constructed by combining the bi-log-concavity constraint with standard confidence bands for $F$ such as, for instance, the Kolmogorov-Smirnov band or \nocite{Owen_1995}{Owen's (1995)} band. A numerical example with the distribution of CEO salaries (\nocite{Woolridge_2000}{Woolridge 2000}) illustrates the usefulness of the proposed method. The benefits of adding the shape constraint are pinpointed in Section~\ref{sec:Consistency}. It is shown that combining a reasonable confidence band with the new shape constraint leads to non-trivial honest confidence bounds for various quantities related to $F$. These include its density, hazard function and reverse hazard function, its moment generating function and arbitrary moments. All proofs are deferred to Section~\ref{sec:Proofs}.

\section{Bi-log-concave distribution functions}
\label{sec:Bi-log-concave}

In what follows we call a distribution function $F$ \textsl{non-degenerate} if the set
\[
	J(F) \ := \ \{x\in\R : 0 < F(x)< 1\}
\]
is nonvoid. Notice that in the case of $J(F) = \emptyset$ the distribution function $F$ would correspond to the Dirac measure $\delta_m$ at some point $m \in \R$, i.e.\ $F(x) = 1_{[x \ge m]}$.

Our first theorem provides three alternative characterisations of bi-log-concavity which are expressed by different constraints for $F$ and its derivatives.

\begin{Theorem}
\label{thm:Characterize}
For a non-degenerate distribution function $F$ the following four statements are equivalent:

\noindent
(i) \ $F$ is bi-log-concave;

\noindent
(ii) \ $F$ is continuous on $\R$ and differentiable on $J(F)$ with derivative $f = F'$ such that
\begin{equation}
\label{eq:Characterize}
	F(x + t) \ \begin{cases}
		\displaystyle
		\le \ F(x) \exp \Bigl( \frac{f(x)}{F(x)} \, t \Bigr) \\
		\displaystyle
		\ge \ 1 - (1 - F(x)) \exp \Bigl( - \, \frac{f(x)}{1 - F(x)} \, t \Bigr)
	\end{cases}
\end{equation}
for arbitrary $x \in J(F)$ and $t \in \R$.

\noindent
(iii) \ $F$ is continuous on $\R$ and differentiable on $J(F)$ with derivative $f = F'$ such that the hazard function $f/(1 - F)$ is non-decreasing and the reverse hazard function $f/F$ is non-increasing on $J(F)$.

\noindent
(iv) \ $F$ is continuous on $\R$ and differentiable on $J(F)$ with bounded and strictly positive derivative $f = F'$. Furthermore, $f$ is locally Lipschitz-continuous on $J(F)$ with $L^1$-derivative $f' = F''$ satisfying
\begin{equation}
\label{eq:Second derivative}
	\frac{-f^2}{1-F}
	\ \le \ f'
	\ \le \ \frac{f^2}{F} .
\end{equation}
\end{Theorem}

The set of all distribution functions $F$ with the properties stated in Theorem~\ref{thm:Characterize} is denoted as $\FFblc$. The inequalities \eqref{eq:Second derivative} in statement~(iv) can be reformulated as follows: $\log f$ is locally Lip\-schitz-continuous on $J(F)$ with $L^1$-derivative $(\log f)'$ satisfying
\[
	(\log(1 - F))' \ \le \ (\log f)' \ \le \ (\log F)' .
\]
(An $L^1$-derivative of a function $h$ on an open interval $J \subset \R$ is a locally integrable function $h'$ on $J$ such that $h(y) - h(x) = \int_x^y h'(t) \, dt$ for all $x,y \in J$.)

\begin{Example}[Bi-modal density]
Consider the mixture $2^{-1} \NN(-\delta,1) + 2^{-1} \NN(\delta,1)$ with $\delta > 0$. Numerical experiments showed that the corresponding c.d.f.\ $F$ is bi-log-concave for $\delta \le 1.34$ but fails to be so for $\delta \ge 1.35$. In case of $\delta = 1.34$, this distribution has a bi-modal density. The corresponding c.d.f.\ $F$ is shown in Figure~\ref{fig:Characterize1}(a), together with the functions $1 + \log F \le F \le - \log(1 - F)$, the inequalities following from $\log(1 + y) \le y$ for arbitrary $y \ge -1$. Bi-log-concavity means that the lower bound $1 + \log F$ is concave while the upper bound $- \log(1 - F)$ is convex. Figures~\ref{fig:Characterize1}-\ref{fig:Characterize2} illustrate the various characterisations of the bi-log-concavity constraint as given in Theorem~\ref{thm:Characterize}. In particular, Figure~\ref{fig:Characterize1}(b) shows the bounds from part~(ii) for one particular point $x \in J(F)$. Figure~\ref{fig:Characterize2}(a) shows the density $f$ together with the hazard function $f/(1 - F)$ and the reverse hazard function $f/F$. It is apparent that the latter two satisfy the monotonicity properties of part~(iii). Figure~\ref{fig:Characterize2}(b) contains the derivative $f'$ together with the bounds $-f^2/(1 - F)$ and $f^2/F$ as given in part~(iv).

\begin{figure}
\centering
(a) $F$ with its concave lower and convex upper bounds.
\includegraphics[width=0.9\textwidth]{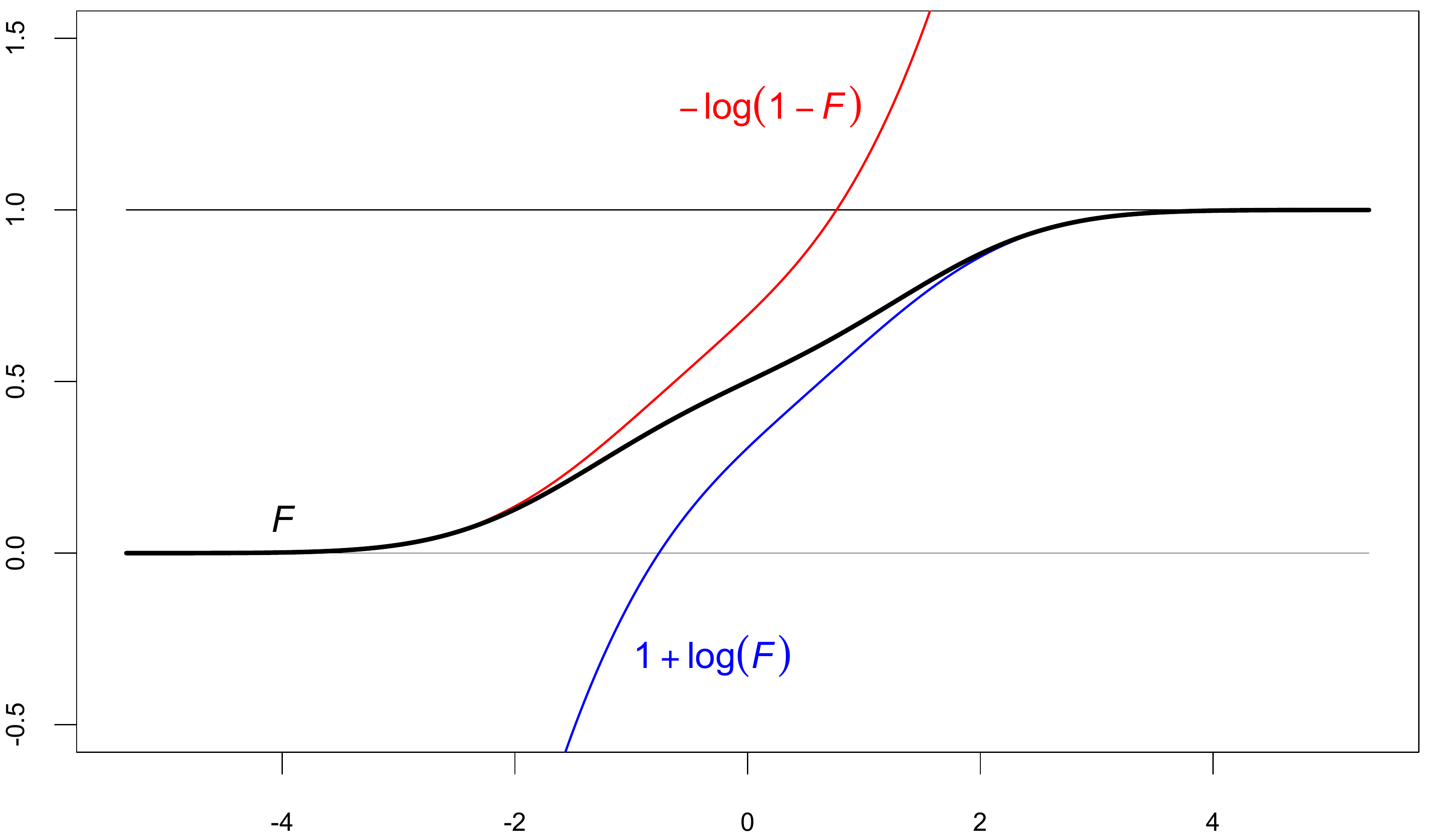}\\
(b) $F$ with the bounds given by Theorem 1 (ii).
\includegraphics[width=0.9\textwidth]{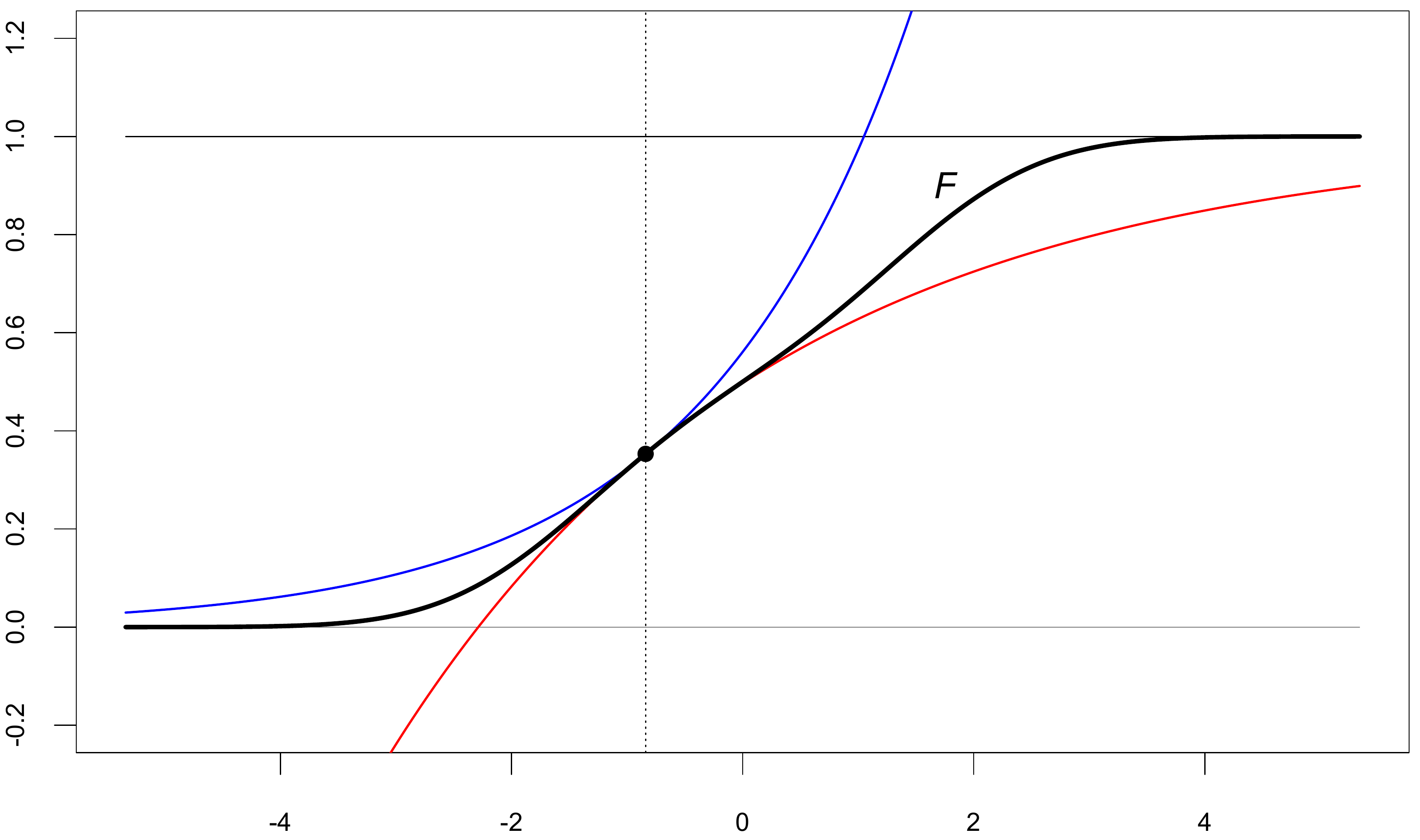}
\caption{A bi-log-concave $F$ with its bounds.}
\label{fig:Characterize1}
\end{figure}

\begin{figure}
\centering
(a) $f$ with monotonic hazard and reversed hazard function as given by Theorem~\ref{thm:Characterize}~(iii).
\includegraphics[width=0.9\textwidth]{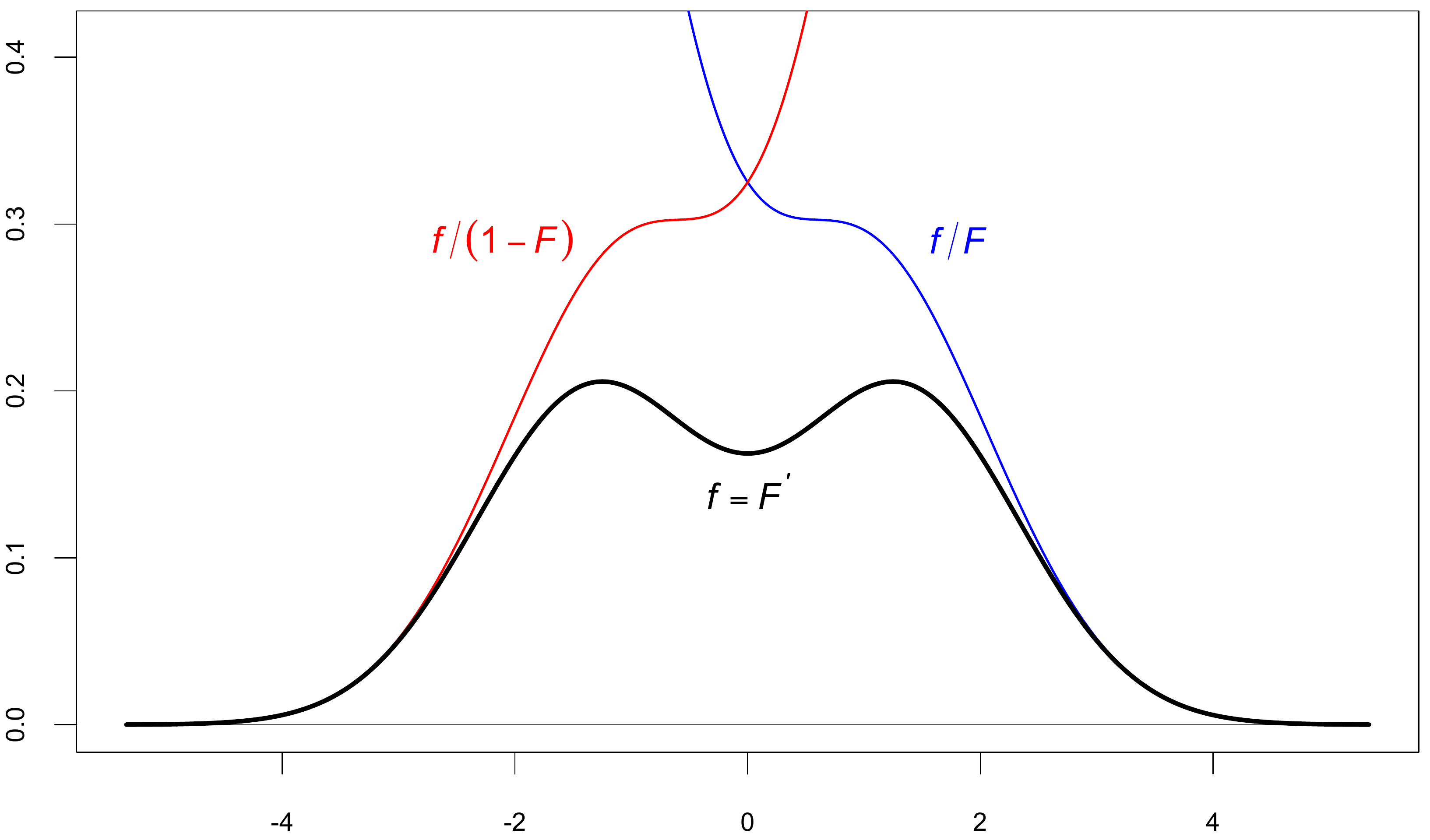}\\
(b) $f'$ with its bounds as given by Theorem~\ref{thm:Characterize}~(iv).
\includegraphics[width=0.9\textwidth]{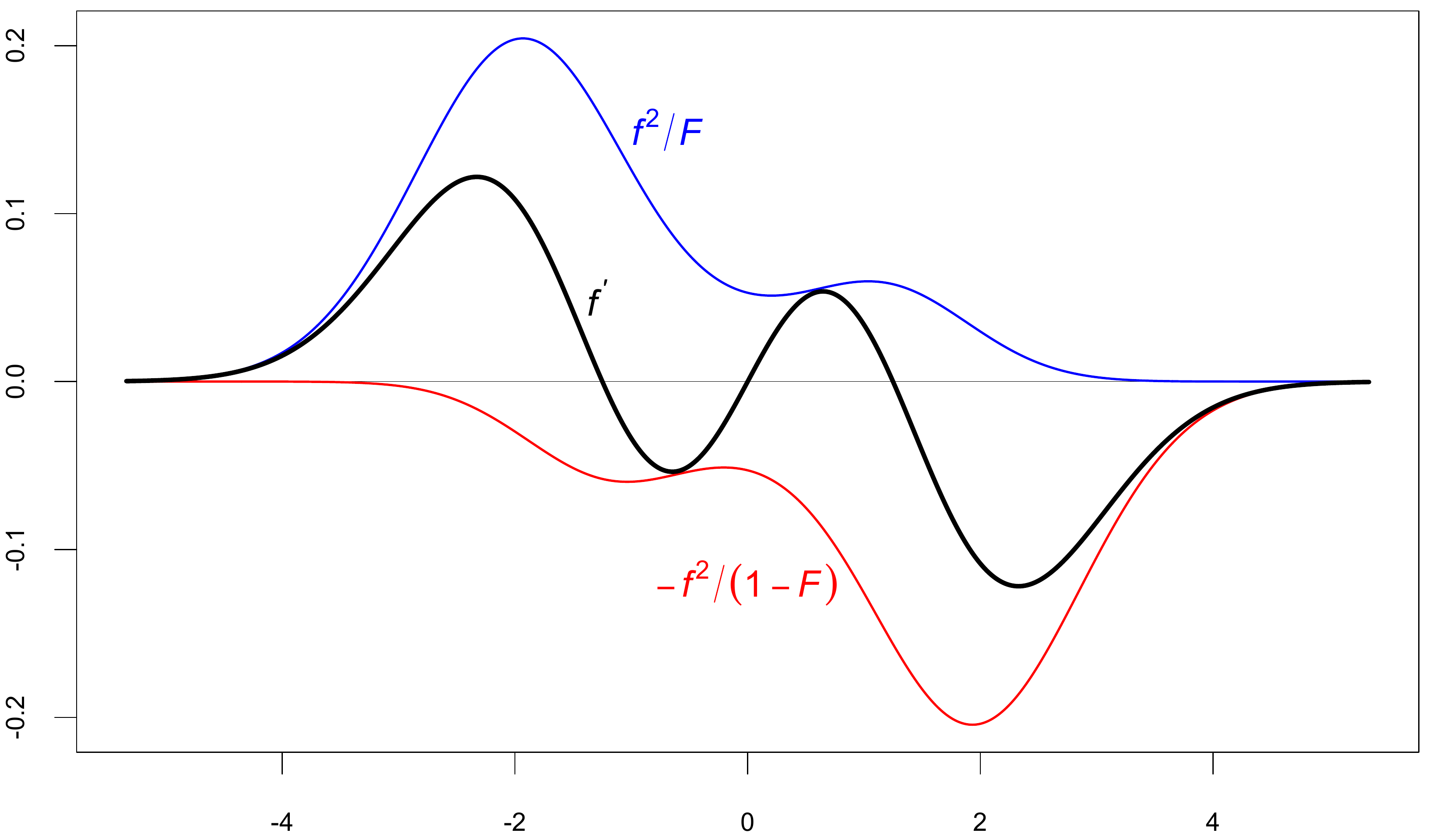}
\caption{Different characterisations of a bi-log-concave $F$.}
\label{fig:Characterize2}
\end{figure}
\end{Example}

While the previous example illustrates bi-log-concavity for a bi-modal density, the next example considers a multi modal density.

\begin{Example}[$k$-modal density] For any integer $k > 0$ and $a \in (0,1)$,
\[
	f(x) \ := \ 1_{[0 < x < 1]} (1 + a \sin(2\pi k x))
\]
defines a probability density with $k$ local maxima. The corresponding c.d.f.\ is given by $F(x) = x + a (1 - \cos(2\pi k x)) / (2\pi k)$ for $x \in [0,1]$, and one can easily deduce from Theorem~\ref{thm:Characterize}~(iv) that it is bi-log-concave if $a$ is sufficiently small.
\end{Example}

\begin{Remark}
For $F \in \FFblc$, its moment-generating function is finite in a neighborhood of $0$. Precisely, it will be shown in Section~\ref{sec:Proofs} that
\begin{equation}
\label{eq:MGF}
	\Bigl\{ t \in \R : \int e_{}^{tx} \, F(dx) < \infty \Bigr\}
	\ = \ \bigl( - T_1(F), T_2(F) \bigr)
\end{equation}
with
\begin{align*}
	T_1(F) \
	&:= \ \sup_{x \in J(F)} \frac{f(x)}{F(x)}
		\ \begin{cases}
			> 0 , \\
			= \infty & \text{if} \ \inf(J(F)) > - \infty ,
		\end{cases} \\
	T_2(F) \
	&:= \ \sup_{x \in J(F)} \frac{f(x)}{1 - F(x)}
		\ \begin{cases}
			> 0 , \\
			= \infty & \text{if} \ \sup(J(F)) < \infty .
		\end{cases}
\end{align*}
\end{Remark}

\section{Confidence bands}
\label{sec:Confidence bands}

A confidence band for $F \in \FFblc$ may be constructed by intersecting a standard confidence band for a (continuous) distribution function with this class $\FFblc$.

\paragraph{Unconstrained nonparametric confidence bands.}
Let $X_1, \ldots, X_n$ be independent random variables with continuous distribution function $F$. In what follows let $(L_n, U_n)$ be a $(1 - \alpha)$-confidence band for $F$ with $0 < \alpha \le 0.5$. This means, $L_n = L_{n,\alpha}(\cdot \,|\, X_1,\ldots,X_n) < 1$ and $U_n = U_{n,\alpha}(\cdot \,|\, X_1,\ldots,X_n) > 0$ are data-driven non-decreasing functions on the real line such that $L_n \le U_n$ pointwise and
\[
	P \bigl( L_n(x) \le F(x) \le U_n(x) \ \text{for all} \ x \in \R \bigr)
	\ = \ 1 - \alpha .
\]

\begin{Example}[Kolmogorov-Smirnov band]
A standard example for $(L_n,U_n)$ is given by
\[
	\bigl[ L_n(x), U_n(x) \bigr]
	\ := \ \Bigl[
			\hat{F}_n(x) - \frac{\kappa_{\alpha,n}^{\rm KS}}{\sqrt{n}} ,
			\hat{F}_n(x) + \frac{\kappa_{\alpha,n}^{\rm KS}}{\sqrt{n}}
		\Bigr] \cap [0,1] ,
\]
where
\[
	\hat{F}_n(x) \ := \ \frac{1}{n} \sum_{i=1}^n 1_{[X_i \le x]} ,
\]
and $\kappa_{n,\alpha}^{\rm KS}$ denotes the $(1 - \alpha)$-quantile of $\sup_{x \in \R} n^{1/2} \bigl| \hat{F}(x) - F(x) \bigr|$; cf.\ \nocite{Shorack_Wellner_1986}{Shorack and Wellner (1986)}. Notice also that $\kappa_{n,\alpha}^{\rm KS} \le \sqrt{\log(2/\alpha)/2}$ by \nocite{Massart_1990}{Massart's (1990)} inequality.
\end{Example}

\begin{Example}[Weighted Kolmogorov-Smirnov band]
Let $X_{(1)} < X_{(2)} < \cdots < X_{(n)}$ denote the order statistics of $X_1,X_2,\ldots,X_n$ and $U_{(i)} := F(X_{(i)})$. It is well known that $U_{(1)} < U_{(2)} < \cdots < U_{(n)}$ are distributed like the order statistics of $n$ independent random variables with uniform distribution on $[0,1]$. By noting that $\Ex(U_{(i)}) = t_i := i/(n+1)$ for $1 \le i \le n$, and using empirical process theory, one can show that for any $\gamma \in [0,1/2)$, the random variable
\begin{equation}
\label{eq:Weighted.KS}
	\sqrt{n} \max_{i=1,2,\ldots,n} \frac{|U_{(i)} - t_i|}{(t_i(1 - t_i))^\gamma}
\end{equation}
converges in distribution to $\sup_{t \in (0,1)} (t(1-t))^{-\gamma}|B(t)| < \infty$ as $n \to \infty$, where $B$ is standard Brownian bridge. In particular, the $(1 - \alpha)$-quantile $\kappa_{n,\alpha}^{\rm WKS}$ of the test statistic \eqref{eq:Weighted.KS} satisfies $\kappa_{n,\alpha}^{\rm WKS} = O(1)$. Inverting this test leads to the $(1 - \alpha)$-confidence band $(L_n,U_n)$ for $F$ with
\[
	\bigl[ L_n(x), U_n(x) \bigr]
	\ = \ \Bigl[
			t_i - \frac{\kappa_{n,\alpha}^{\rm WKS}}{\sqrt{n}}
				(t_i(1 - t_i))^\gamma,
			t_{i+1} + \frac{\kappa_{n,\alpha}^{\rm WKS}}{\sqrt{n}}
				(t_{i+1}(1 - t_{i+1}))^\gamma
		\Bigr] \cap [0,1]
\]
for $i \in \{0,1,\ldots,n\}$ and $x \in [X_{(i)}, X_{(i+1)})$. Here $X_{(0)} := -\infty$ and $X_{(n+1)} := \infty$.
\end{Example}

\begin{Example}[Owen's band refined]
\label{ex:Owen.refined}
Another confidence band which may be viewed as a refinement of \nocite{Owen_1995}{Owen's (1995)} method has been proposed recently by \nocite{Duembgen_Wellner_2014}{D\"umbgen and Wellner (2014)}. Let
\[
	K(\hat{p},p) \ := \ \hat{p} \log \frac{\hat{p}}{p}
		+ (1 - \hat{p}) \log \frac{1 - \hat{p}}{1 - p}
\]
for $p, \hat{p} \in [0,1]$ with the usual conventions that $0 \log(\cdot) := 0$ and $a \log(a/0) := \infty$ for $a > 0$. Furthermore, for $t \in (0,1)$ let
\[
	C(t) \ := \ \log(1 + \mathrm{logit}(t)^2/2) / 2
	\quad\text{and}\quad
	D(t) \ := \ \log(1 + C(t)^2/2) / 2 .
\]
Then for any fixed $\nu > 2$,
\begin{equation}
\label{eq:ODW}
	\max_{j = 1,2,\ldots,n} \,
	\bigl( (n+1) K(t_j, U_{(j)}) - C(t_j) - \nu D(t_j) \bigr)
\end{equation}
converges in distribution to
\[
	\sup_{t \in (0,1)} \Bigl( \frac{B(t)^2}{t(1 - t)} - C(t) - \nu D(t) \Bigr)
	\ < \ \infty .
\]
In particular, the $(1 - \alpha)$-quantile $\kappa_{n,\alpha}^{\rm ODW}$ of the test statistic \eqref{eq:ODW} is bounded as $n \to \infty$. Inverting this test leads to the following confidence band $(L_n,U_n)$:
\begin{align*}
	L_n(x) \
	&:= \ 0
		\quad \text{for} \ x < X_{(1)} , \\
	L_n(x) \
	&:= \ \min \bigl\{ p \in (0,t_j] : K(t_j,p) \le \gamma_n(t_j) \bigr\}
		\quad\text{for} \ 1 \le j \le n, X_{(j)} \le x < X_{(j+1)} , \\
	U_n(x) \
	&:= \ \max \bigl\{ p \in [t_j,1) : K(t_j,p) \le \gamma_n(t_j) \bigr\}
		\quad\text{for} \ 1 \le j \le n, X_{(j-1)} \le x < X_{(j)} , \\
	U_n(x) \
	&:= \ 1
		\quad \text{for} \ x \ge X_{(n)} ,
\end{align*}
where
\[
	\gamma_n(t)
	\ := \ \frac{C(t) + \nu D(t) + \kappa_{n,\alpha}^{\rm ODW}}{n+1} .
\]
\end{Example}

\paragraph{Confidence bands for a bi-log-concave F.}
Now suppose that $F$ belongs to $\FFblc$. Under this assumption, a $(1 - \alpha)$-confidence band $(L_n,U_n)$ for $F$ may be refined as follows:
\begin{align*}
	L_n^o(x) \
	&:= \ \inf \bigl\{ G(x) : G \in \FFblc, L_n \le G \le U_n \bigr\} , \\
	U_n^o(x) \
	&:= \ \sup \bigl\{ G(x) : G \in \FFblc, L_n \le G \le U_n \bigr\} .
\end{align*}
It may happen that no bi-log-concave distribution function fits into the band $(L_n,U_n)$. In this case we set $L_n^o \equiv 1$ and $U_n^o \equiv 0$ and conclude with confidence $1 - \alpha$ that $F \not\in \FFblc$. But in the case of $F \in \FFblc$ this happens with probability at most $\alpha$. Indeed, the construction of $(L_n^o,U_n^o)$ implies that
\[
	\Pr(L_n^o \le F \le U_n^o) \ = \ \Pr(L_n \le F \le U_n)
	\quad\text{if} \ F \in \FFblc .
\]

The following algorithm is used to determine the refined band $(L_n^o,U_n^o)$. An essential ingredient is a procedure $\mathrm{ConcInt}(\cdot,\cdot)$ (concave interior). Given any finite set $\mathcal{T} = \{t_0, t_1, \ldots, t_m\}$ of real numbers $t_0 < t_1 < \cdots < t_m$ and any pair $(\ell,u)$ of functions $\ell, u : \mathcal{T} \to [-\infty,\infty)$ with $\ell < u$ pointwise and $\ell(t) > -\infty$ for at least two different points $t \in \mathcal{T}$, this procedure computes the pair $(\ell^o,u^o)$, where
\begin{align*}
	\ell^o(x) \
	&:= \ \inf \bigl\{ g(x) : g \ \text{concave on} \ \R, \
		\ell \le g \le u \ \text{on} \ \mathcal{T} \bigr\} , \\
	u^o(x) \
	&:= \ \sup \bigl\{ g(x) : g \ \text{concave on} \ \R, \
		\ell \le g \le u \ \text{on} \ \mathcal{T} \bigr\} .
\end{align*}
This is a standard and solvable problem. On the one hand, $\ell^o$ is the smallest concave majorant of $\ell$ on $\mathcal{T}$ which may be computed via a suitable version of the pool-adjacent-violators algorithm (\nocite{Robertson_etal_1988}{Robertson et al.\ 1988}). Indeed, there exist indices $0 \le j(0) < j(1) < \cdots < j(b) \le m$ such that
\[
	\ell^o \ \begin{cases}
		\equiv \ - \infty \quad \text{on} \ \R \setminus [t_{j(0)}, t_{j(b)}] , \\
		\text{is linear on} \ [t_{j(a-1)}, t_{j(a)}] \ \text{for} \ 1 \le a \le b , \\
		\text{changes slope at} \ t_{j(a)} \ \text{if} \ 1 \le a < b .
	\end{cases}
\]
Having computed $\ell^o$, we can check whether $\ell^o \le u$ on $\mathcal{T}$. If this is not the case, there is no concave function fitting in between $\ell$ and $u$, and the procedure returns a corresponding error message. Otherwise the value of $u^o(x)$ equals
\[
	\min \Bigl\{ u(s) + \frac{u(s) - \ell^o(r)}{s - r} (x - s) :
		r \in \mathcal{T}_o, s \in \mathcal{T}, \
		r < s \le x \ \text{or} \ x \le s < r \Bigr\} ,
\]
where $\mathcal{T}_o = \{t_{j(0)}, t_{j(1)}, \ldots, t_{j(b)}\}$. To maximise $g(x)$ over all concave functions $g$ such that $\ell \le g \le u$, we may assume without loss of generality that for fixed $x$ and a given value $y$ of $g(x)$, the function $g$ is the smallest concave function such that $g \ge \ell_o$ and $g(x) = y$. But the latter function is piecewise linear with changes of slope at $x$ and some points in $\mathcal{T}_o$. Moreover, if $y$ is chosen as large as possible, $g(s)$ has to be equal to $u(s)$ for at least one point $s \in \mathcal{T}$.

\begin{figure}
\centering
\includegraphics[width=0.95\textwidth]{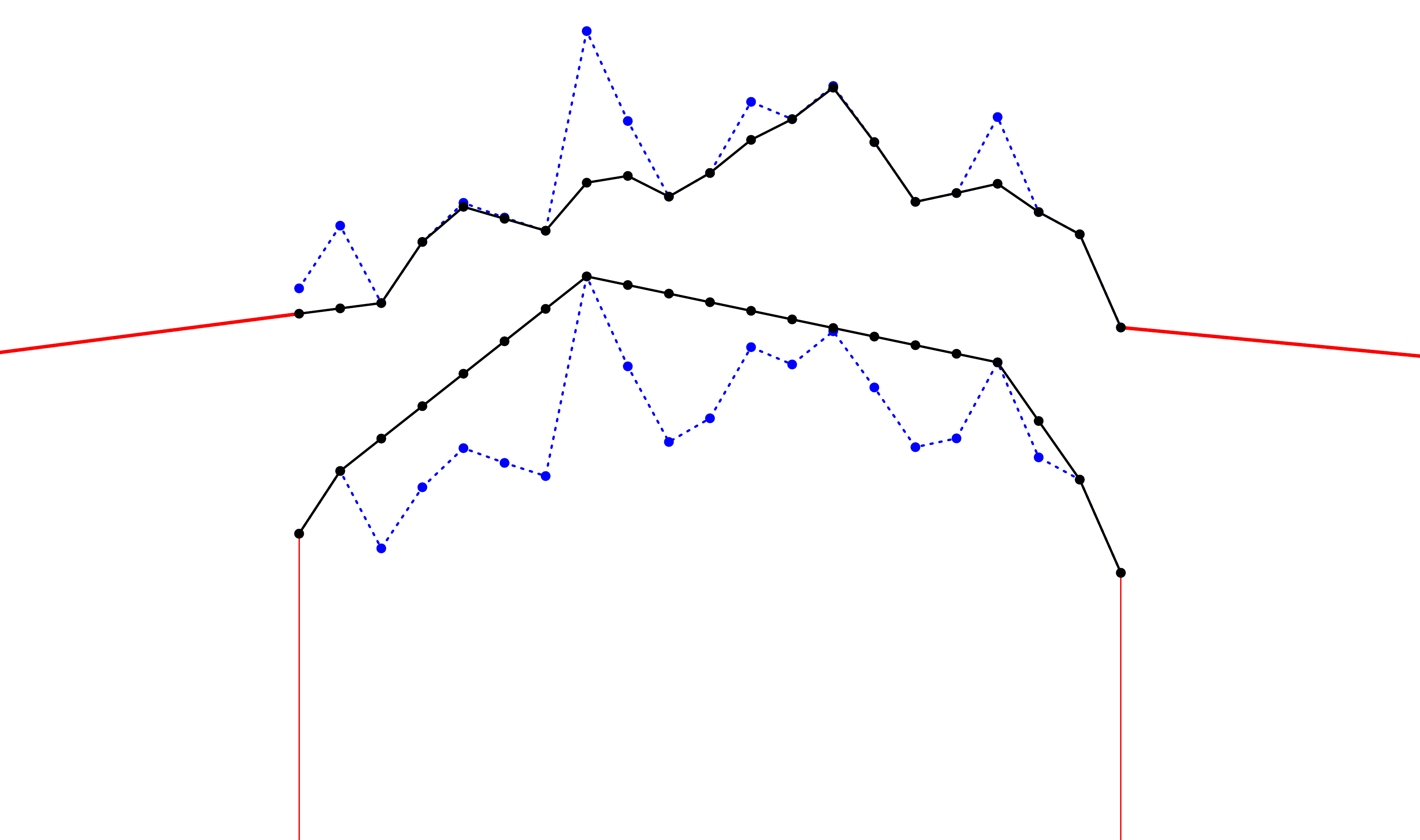}
\caption{Graphical illustration of the procedure $\mathrm{ConcInt}(\cdot,\cdot)$.}
\label{fig:ConcaveInterior}
\end{figure}

Figure~\ref{fig:ConcaveInterior} illustrates this procedure for $\mathcal{T}$ consisting of 21 points. It shows two (parallel) functions $\ell$ and $u$ evaluated at all points in $\mathcal{T}$, indicated by bullets and interpolating dashed lines. In addition the plot shows the resulting functions $\ell^o$ and $u^o$ on $\mathcal{T} \cup (-\infty,t_0) \cup (t_m,\infty)$, which are displayed as interpolating solid lines.

In our context, $\mathcal{T}$ is chosen as a fine grid of points such that $t_0 < X_{(1)}$ and $t_m > X_{(n)}$ and $\{X_1,X_2,\ldots,X_n\} \subset \mathcal{T}$. Table~\ref{tab:code} contains pseudo-code for our algorithm to compute $(L_n^o, U_n^o)$. We tacitly assume that whenever $\mathrm{ConcInt}(\cdot,\cdot)$ returns an error message, the whole algorithm stops and reports the fact that there is no $G \in \FFblc$ satisfying $L_n \le G \le U_n$.

\begin{table}
\[
	\begin{array}{|l|}
	\hline
	(L_n^o, U_n^o) \leftarrow
		(L_n,U_n) \\[0.5ex]
	(\ell_o,u_o) \leftarrow
		\mathrm{ConcInt} \bigl( \log(L_n^o), \log(U_n^o) \bigr) \\
	(\tilde{L}_n^o,\tilde{U}_n^o) \leftarrow
		\bigl( \exp(\ell^o), \exp(u^o) \bigr) \\[0.5ex]
	(\ell_o,u_o) \leftarrow
		\mathrm{ConcInt} \bigl( \log(1 - \tilde{U}_n^o),
			\log(1 - \tilde{L}_n^o) \bigr) \\
	(\tilde{L}_n^o,\tilde{U}_n^o) \leftarrow
		\bigl( 1 - \exp(u^o), 1 - \exp(\ell^o) \bigr) \\[1ex]
	\text{while} \
		(\tilde{L}_n^o,\tilde{U}_n^o) \ne (L_n^o,U_n^o) \ \text{do} \\[0.5ex]
	\strut\quad
		(L_n^o, U_n^o) \leftarrow
			(\tilde{L}_n^o,\tilde{U}_n^o) \\[0.5ex]
	\strut\quad
		(\ell_o,u_o) \leftarrow
			\mathrm{ConcInt} \bigl( \log(L_n^o), \log(U_n^o) \bigr) \\
	\strut\quad
		(\tilde{L}_n^o,\tilde{U}_n^o) \leftarrow
			\bigl( \exp(\ell^o), \exp(u^o) \bigr) \\[0.5ex]
	\strut\quad
		(\ell_o,u_o) \leftarrow
			\mathrm{ConcInt} \bigl( \log(1 - \tilde{U}_n^o),
				\log(1 - \tilde{L}_n^o) \bigr) \\
	\strut\quad
		(\tilde{L}_n^o,\tilde{U}_n^o) \leftarrow
			\bigl( 1 - \exp(u^o), 1 - \exp(\ell^o) \bigr) \\[0.5ex]
	\text{end while} \\
	\hline
	\end{array}
\]
\caption{Pseudocode for the computation of $(L_n^o,U_n^o)$.}
\label{tab:code}
\end{table}

The next lemma implies that our proposed new band $(L_n^o,U_n^o)$ has some desirable properties under rather weak conditions on $(L_n,U_n)$. In particular, both $L_n^o$ and $U_n^o$ are Lipschitz-continuous on $\R$, unless $\inf\{x \in \R : L_n(x) > 0\} \ge \sup\{x \in \R : U_n(x) < 1\}$. Moreover, if $\lim_{x \to \infty} L_n(x) > \lim_{x \to - \infty} U_n(x)$, then $U_n^o(x)$ converges exponentially fast to $0$ as $x \to -\infty$ while $L_n^o(x)$ converges exponentially fast to $1$ as $x \to \infty$.

\begin{Lemma}
\label{lem:continuity.exponential.tails}
For real numbers $a < b$ and $0 < r < s < 1$ define
\[
	\gamma_1 \ := \ \frac{\log(s/r)}{b - a}
	\quad\text{and}\quad
	\gamma_2 \ := \ \frac{\log \bigl( (1 - r)/(1 - s) \bigr)}{b - a} .
\]

\noindent
\textbf{(i)} \ If $L_n(a) \ge r$ and $U_n(b) \le s$, then $L_n^o$ and $U_n^o$ are Lipschitz-continuous on $\R$ with Lipschitz constant $\max\{\gamma_1, \gamma_2\}$.

\noindent
\textbf{(ii)} \ If $U_n(a) \le r$ and $L_n(b) \ge s$, then
\begin{align*}
	U_n^o(x) \
	&\le \ r \exp(\gamma_1 (x - a))
	\quad\text{for} \ x \le a
\intertext{and}
	1 - L_n^o(x) \
	&\le \ (1 - s) \exp(- \gamma_2 (x - b))
	\quad\text{for} \ x \ge b .
\end{align*}
\end{Lemma}

\subsection{A numerical example}

We illustrate our methods with a data set from \nocite{Woolridge_2000}{Woolridge (2000)}. It contains for $n = 177$ randomly chosen companies in the U.S.\ the annual salaries of their CEOs in 1990, rounded to multiples of $1000$ USD. Since it is not clear to us how the rounding has been done, we assume that an observation $Y_{i,{\rm raw}} \in \mathbb{N}$ corresponds to an unobserved true salary $Y_i$ within $(Y_{i,{\rm raw}}-1, Y_{i,{\rm raw}} + 1)$, and we consider $Y_1, Y_2, \ldots, Y_n$ to be a random sample from a distribution function $G$ on $(0,\infty)$. Salary distributions are well-known to be heavily right-skewed with heavy right tails. A standard model is that $Y \sim G$ has the same distribution as $10^X$ for some Gaussian random variable $X$, see \nocite{Kleiber_Kotz_2003}{Kleiber and Kotz (2003)}. We assume that the distribution function $F(x) := G(10^x)$ of $X_i := \log_{10}(Y_i)$ is bi-log-concave. More specifically, we compute an unrestricted confidence band $(L_n,U_n)$, where $L_n$ is computed with $\bigl( \log_{10}(Y_{i,{\rm raw}} + 1) \bigr)_{i=1}^n$ and $U_n$ with $\bigl( \log_{10}(Y_{i,{\rm raw}} - 1) \bigr)_{i=1}^n$.

\begin{figure}
\centering
(a) Kolmogorov-Smirnov confidence bands:
\includegraphics[width=0.99\textwidth]{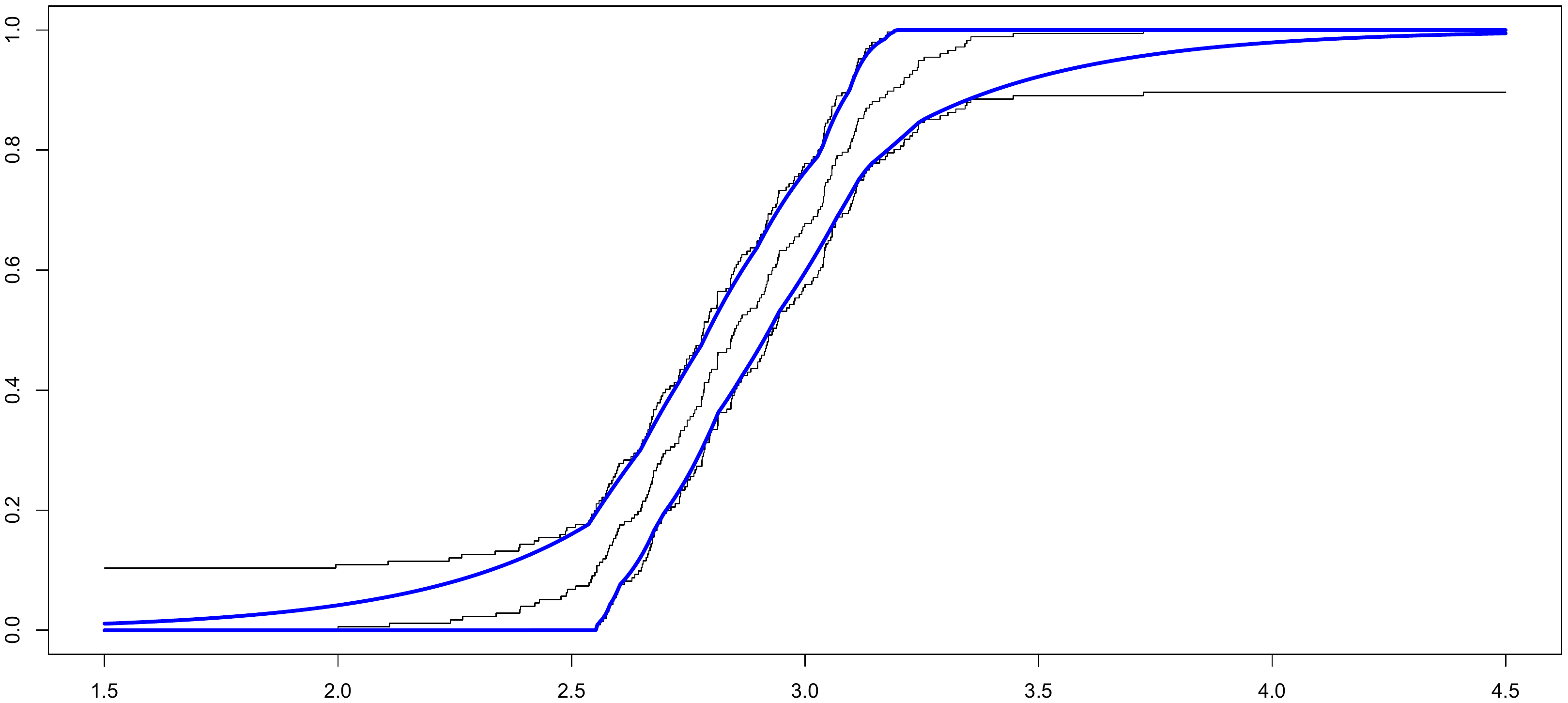}
\centering
(b) Weighted Kolmogorov-Smirnov bands:
\includegraphics[width=0.99\textwidth]{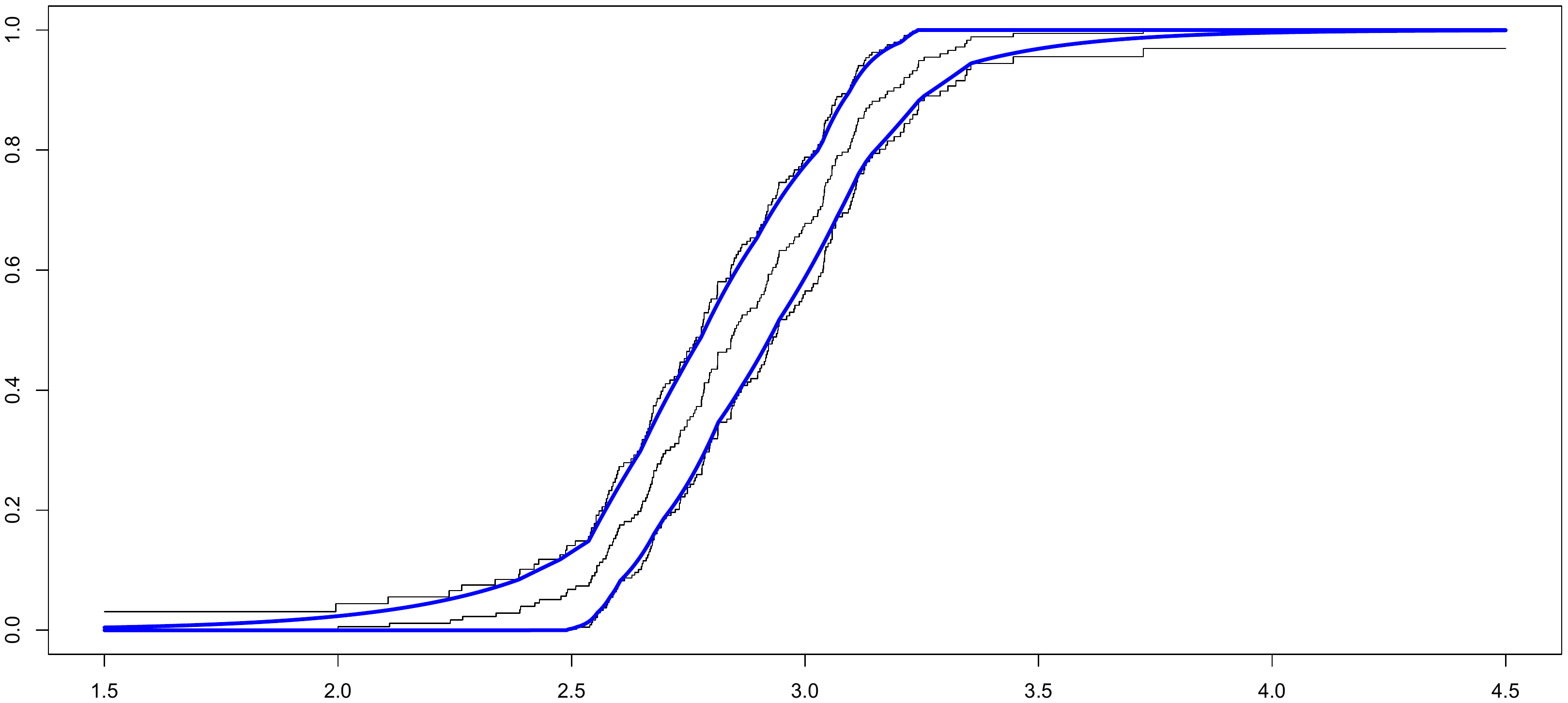}
\caption{Estimated distribution function with unconstrained and constrained confidence bands for CEO salaries.}
\label{fig:CEOSAL2}
\end{figure}

Figure~\ref{fig:CEOSAL2}(a) shows the Kolmogorov-Smirnov $95\%$-confidence bands for $F$, without (black lines) and with (blue lines) the restriction of bi-log-concavity. Figure~\ref{fig:CEOSAL2}(b) shows the confidence bands based on the weighted Kolmogorov-Smirnov $95\%$-confidence band, where $\gamma = 0.4$. The corresponding quantiles have been estimated in $2\cdot 10^6$ Monte Carlo simulations. In both cases the shape constraint yields a substantial gain of precision. Notice also that the bounds in Figure~\ref{fig:CEOSAL2}(b) are tighter in the tails but slightly wider in the central part than those in Figure~\ref{fig:CEOSAL2}(a), for the unconstrained band as well as for the band with shape constraint.

\section{Consistency properties}
\label{sec:Consistency}

In this section we study the asymptotic behaviour of the proposed confidence band $(L_n^o, U_n^o)$ when $F \in \FFblc$. Or goal is to pinpoint the benefits of utilizing the shape constraint of bi-log-concavity. All asymptotic statements refer to $n \to \infty$ while $F$ is fixed.

We start with rather general consistency results for $(L_n^o,U_n^o)$. Recall that we set $L_n^o \equiv 1$ and $U_n^o \equiv 0$ in the case of no $G \in \FFblc$ fitting in between $L_n$ and $U_n$, concluding with confidence $1 - \alpha$ that $F \not\in \FFblc$. The supremum norm of a function $h : \R \to \R$ is denoted by $\|h\|_\infty = \sup_{x \in \R} |h(x)|$, and for $K \subset \R$ we write $\|h\|_{K,\infty} := \sup_{x \in K} |h(x)|$.

\begin{Theorem}
\label{thm:consistency}
Suppose that the original confidence band $(L_n,U_n)$ is consistent in the sense that for any fixed $x \in \R$, both $L_n(x)$ and $U_n(x)$ tend to $F(x)$ in probability.

\noindent
\textbf{(i)} \ Suppose that $F \not\in \FFblc$. Then $\Pr(L_n^o \le U_n^o) \to 0$.

\noindent
\textbf{(ii)} \ Suppose that $F \in \FFblc$. Then $\Pr(L_n^o \le U_n^o) \ge 1 - \alpha$, and
\[
	\sup_{G \in \FFblc \,:\, L_n \le G \le U_n} \, \|G - F\|_\infty
	\ \to_p \ 0 ,
\]
where $\sup(\emptyset) := 0$. Moreover, for any compact interval $K \subset J(F)$,
\[
	\sup_{G \in \FFblc \,:\, L_n \le G \le U_n} \, \|h_G - h_F\|_{K,\infty}
	\ \to_p \ 0 ,
\]
where $h_G$ stands for any of the three functions $G'$, $\log(G)'$ and $\log(1 - G)'$. Finally, for any fixed $x_1 \in J(F)$ and $b_1 < f(x_1)/F(x_1)$,
\[
	\Pr \bigl( U_n^o(x) \le U_n(x') \exp(b_1(x - x'))
		\ \text{for} \ x \le x' \le x_1 \bigr)
	\ \to \ 1 ,
\]
while for any fixed $x_2 \in J(F)$ and $b_2 < f(x_2)/(1 - F(x_2))$,
\[
	\Pr \bigl( 1 - L_n^o(x) \le (1 - L_n(x')) \exp(- b_2(x - x'))
		\ \text{for} \ x \ge x' \ge x_2 \bigr)
	\ \to \ 1 .
\]
\end{Theorem}

A direct consequence of Theorem~\ref{thm:consistency} are consistent confidence bounds for functionals $\int \phi \, dF$ of $F$ with well-behaved integrands $\phi : \R \to \R$:

\begin{Corollary}
\label{cor:consistency}
Suppose that the original confidence band $(L_n,U_n)$ is consistent, and let $F \in \FFblc$. Let $\phi : \R \to \R$ be absolutely continuous with a derivative $\phi'$ satisfying the following constraint: For constants $a \in \R$ and $0 \le b_1 < T_1(F)$, $0 \le b_2 < T_2(F)$,
\[
	|\phi'(x)| \le \ \exp(a + b_1 x^- + b_2 x^+)
\]
with $x^\pm := \max\{\pm x, 0\}$. Then
\[
	\sup_{G \,:\, L_n^o \le G \le U_n^o}
		\Bigl| \int \phi \, dG - \int \phi \, dF \Bigr|
	\ \to_p \ 0 .
\]
\end{Corollary}

The previous supremum is meant over all distribution functions $G$ within the confidence band $(L_n^o,U_n^o)$, which is larger than the supremum over all distribution funcions $G \in \FFblc$ between $L_n$ and $U_n$. Corollary~\ref{cor:consistency} applies to $\phi(x) := e^{tx}$ with $- T_1(F) < t < T_2(F)$. Indeed, the proof of \eqref{eq:MGF} implies the following explicit formulae in the case $L_n^o \le U_n^o$:
\begin{align*}
	\inf_{G \,:\, L_n^o \le G \le U_n^o} \int e^{tx} G(dx) \
	&= \ \begin{cases}
		\displaystyle
		\int_{\R} t e^{tx} (1 - U_n^o(x)) \, dx
		& \text{if} \ t > 0 , \\[2ex]
		\displaystyle
		\int_{\R} |t| e^{tx} L_n^o(x) \, dx
		& \text{if} \ t < 0 ,
		\end{cases} \\
	\sup_{G \,:\, L_n^o \le G \le U_n^o} \int e^{tx} \, G(dx) \
	&= \ \begin{cases}
		\displaystyle
		\int_{\R} t e^{tx} (1 - L_n^o(x)) \, dx
		&\text{if} \ t > 0 , \\[2ex]
		\displaystyle
		\int_{\R} |t| e^{tx} U_n^o(x) \, dx
		&\text{if} \ t < 0 .
		\end{cases}
\end{align*}

Now we refine Corollary~\ref{cor:consistency} by providing rates of convergence, assuming that the original confidence band $(L_n,U_n)$ satisfies the following property:

\paragraph{Condition (*)}
For certain constants $\gamma \in [0,1/2)$ and $\kappa, \lambda > 0$,
\[
	\max\{\hat{F}_n - L_n, U_n - \hat{F}_n\}
			\ \le \ \kappa n^{-1/2} (\hat{F}_n (1 - \hat{F}_n))^\gamma
\]
on the interval $\{\lambda n^{-1/(2 - 2\gamma)} \le \hat{F}_n \le 1 - \lambda n^{-1/(2 - 2\gamma)}\}$.

Obviously this condition is satisfied with $\gamma = 0$ in the case of the Kolmogorov-Smirnov band. For the weighted Kolmogorov-Smirnov band it is satisfied with the given value of $\gamma \in [0,1/2)$. In the refined version of Owen's band, it is satisfied for \textsl{any} fixed number $\gamma \in (0,1/2)$.

\begin{Theorem}
\label{thm:consistency.rates}
Suppose that $F \in \FFblc$, and let $(L_n,U_n)$ satisfy Condition~(*). Let $\phi : \R \to \R$ be absolutely continuous.

\noindent
\textbf{(i)} \ Suppose that $|\phi'(x)| = O(|x|^{k-1})$ as $|x| \to \infty$ for some number $k \ge 1$. Then
\[
	\sup_{G \,:\, L_n^o \le G \le U_n^o}
		\Bigl| \int \phi \, dG - \int \phi \, dF \Bigr|
	\ = \ \begin{cases}
		O_p \bigl( n^{-1/2} (\log n)^k \bigr)
			& \text{if} \ \gamma = 0 , \\
		O_p(n^{-1/2})
			& \text{if} \ \gamma > 0 .
	\end{cases}
\]

\noindent
\textbf{(ii)} \ Suppose that $\phi$ satisfies the conditions in Corollary~\ref{cor:consistency}. Then
\begin{equation}
\label{eq:rate.mgf}
	\sup_{G \,:\, L_n^o \le G \le U_n^o}
		\Bigl| \int \phi \, dG - \int \phi \, dF \Bigr|
	\ = \ O_p(n^{-\beta})
\end{equation}
for any exponent $\beta \in (0,1/2]$ such that
\[
	\beta
	\ < \ \frac{1 - \max \bigl\{ b_1/T_1(F), b_2/T_2(F) \bigr\}}
		{2(1 - \gamma)} .
\]
\end{Theorem}

The additional factor $(\log n)^k$ in part~(i) cannot be avoided. To verify this we consider $\phi(x) = x^k$ and the distribution function $F$ of a standard exponential random variable $X$, i.e.\ $F(x) = 1 - e^{-x}$ for $x \ge 0$. Further let $F_n$ be the conditional distribution function of $X$, given that $X \le x_n := (\log n)/2 - \log c$ with a fixed $c > 0$. Then both $F$ and $F_n$ are bi-log-concave, $\|F_n - F\|_\infty = e^{-x_n} = c n^{-1/2}$, but
\begin{align*}
	\int \phi \, d(F_n - F) \
	&= \ \Ex(X^k) - \Ex(X^k \,|\, X \le x_n) \\
	&= \ \Pr(X > x_n)
		\bigl( \Ex(X^k \,|\, X > x_n) - \Ex(X^k \,|\, X \le x_n) \bigr) \\
	&\ge \ \Pr(X > x_n) \bigl( x_n^k - \Ex(X^k) / \Pr(X \le x_n) \bigr) \\
	&= \ 2^{-k} c n^{-1/2} (\log n)^k (1 + o(1)) .
\end{align*}
Consequently, if we use the Kolmogorov-Smirnov confidence band, the asymptotic probability of $n^{1/2} \|\hat{F}_n - F\|_\infty \le \kappa_{n,\alpha}^{\rm KS} - c$ is strictly positive, provided that $0 < c < \lim_{n \to \infty} \kappa_{n,\alpha}^{\rm KS}$. But then $F_n$ satisfies $n^{1/2} \|F_n - \hat{F}_n\|_\infty \le \kappa_{n,\alpha}^{\rm KS}$, so $L_n^o \le F_n \le U_n^o$, and the $k$-th moments of $F$ and $F_n$ differ by $2^{-k} c n^{-1/2} (\log n)^k (1 + o(1))$.

If $(L_n^o,U_n^o)$ is constructed with the refined version of Owen's confidence band, we may choose $\gamma$ arbitrarily close to $1/2$, so the term $2(1 - \gamma)$ is arbitrarily close to $1$. Thus \eqref{eq:rate.mgf} holds for any exponent $\beta \in (0,1/2]$ such that
\[
	\beta
	\ < \ 1 - \max \bigl\{ b_1/T_1(F), b_2/T_2(F) \bigr\} .
\]
In particular,
\[
	\sup_{G \,:\, L_n^o \le G \le U_n^o}
		\Bigl| \int e^{tx} \, G(dx) - \int e^{tx} \, F(dx) \Bigr|
	\ = \ O_p(n^{-1/2})
\]
whenever $- T_1(F)/2 < t < T_2(F)/2$.

\section{Proofs}
\label{sec:Proofs}

When proving Theorem~\ref{thm:Characterize} we assume that the reader is acquainted with the following facts about concave functions:

\begin{Lemma}
\label{lem:Concavity}
Suppose that $h:\R\rightarrow[-\infty,+\infty)$ is a concave function. Then it satisfies the following properties:\\[1ex]
(i) \ $h$ is continuous on the interior of $\{h> -\infty\}:=\{x\in\R:\textrm{ }h(x)>-\infty\}$.\\[0.5ex]
(ii) \ For each interior point $x$ of $\{h> -\infty\}$, the left- and right-sided derivatives $h'(x\,-)$ and $h'(x\,+)$ exist in $\R$ and satisfy $h'(x\,-) \ge h'(x\,+)$. Moreover, $h(x\,\pm)$ is non-decreasing in $x$.\\[0.5ex]
(iii) \ For each interior point $x$ of $\{h> -\infty\}$ and $a \in [h'(x\,+), h'(x\,-)]$,
\[
	h(x + t) \ \le \ h(x) + at
	\quad\text{for all} \ t \in \R .
\]
\end{Lemma}

Here is a second useful result:

\begin{Lemma}
\label{lem:Elementary}
Let $h$ be a real-valued function on an open interval $J \subset \R$, and let $[a,b] \in [-\infty,\infty]$. Then the following two statements are equivalent:

\noindent
(i) \ For arbitrary different $x,y \in J$,
\[
	\frac{h(y) - h(x)}{y - x} \ \in \ [a,b] .
\]

\noindent
(ii) \ For arbitrary $x \in J$,
\[
	\liminf_{y \to x}
		\frac{h(y) - h(x)}{y - x} \ \ge \ a
	\quad\text{and}\quad
	\limsup_{y \to x}
		\frac{h(y) - h(x)}{y - x} \ \le \ b .
\]
\end{Lemma}

In the case of $[a,b] = [0,\infty]$ or $[a,b] = [-\infty,0]$, part~(i) is equivalent to $h$ being non-decreasing or non-increasing, respectively. In the case of $[a,b] \subset \R$, part~(i) is equivalent to $h$ having an $L^1$-derivative $h'$ on $J$ with values in $[a,b]$.

Lemma~\ref{lem:Elementary} follows essentially from a bisection argument and the following observation: For points $r < s < t$ in $J$,
\[
	\frac{h(t) - h(r)}{t - r}
	\ = \ \alpha \, \frac{h(s) - h(r)}{s - r} + (1 - \alpha) \, \frac{h(t) - h(s)}{t - s}
\]
with $\alpha := (s - r)/(t - r) \in (0,1)$. In particular,
\[
	\frac{h(t) - h(r)}{t - r} \
	\begin{cases}
		\displaystyle
		\ge \ \min \Bigl\{ \frac{h(s) - h(r)}{s - r}, \frac{h(t) - h(s)}{t - s} \Bigr\} ,
		\\[1.5ex]
		\displaystyle
		\le \ \max \Bigl\{ \frac{h(s) - h(r)}{s - r}, \frac{h(t) - h(s)}{t - s} \Bigr\} .
	\end{cases}
\]

\begin{proof}[\bf Proof of Theorem~\ref{thm:Characterize}]
Equivalence of (i-iv) will be verified in four steps.

\noindent
\textbf{Proof of (i) $\Rightarrow$ (ii).} \
Suppose that $F$ is bi-log-concave. Since $\log F$ is concave, it follows from Lemma~\ref{lem:Concavity} that $F$ is continuous on $(a,\infty)$, where $a := \inf \{F > 0\}$. Furthermore, concavity of $\log(1 - F)$ implies that $F$ is continuous on $(-\infty,b)$ with $b := \sup \{F < 1\} \ge a$. But $a < b$, because otherwise $F$ would be degenerate. Hence $F$ is continuous on $\R$. In particular, $J(F)$ is the open and nonvoid interval $(a,b)$.

Concavity of $h := \log F$ implies that for $a < x < b$ its left- and right-sided derivatives $h'(x\,-), h'(x\,+)$ exist in $\R$ and satisfy $h'(x \, -) \ge h'(x \, +)$. But then
\[
	F'(x\,\pm)
	\ = \ \lim_{t \to 0, \, \pm t > 0} \frac{\exp(h(x+t)) - \exp(h(x))}{t}
	\ = \ F(x) h'(x\,\pm)
\]
exist in $\R$, too, and satisfy the inequalities
\[
	F'(x\,-) \ \ge \ F'(x\,+) .
\]
Analogously one can deduce from concavity of $h := \log(1 - F)$ that
\[
	- F'(x\,-) = (1 - F)'(x\,-) \ \ge \ (1 - F)'(x\,+) = - F'(x\,+) ,
\]
so that $F'(x\,-) = F'(x\,+)$. This proves differentiability of $F$ on $J(F)$.

Finally, the inequalities \eqref{eq:Characterize} follow directly from the last part of Lemma~\ref{lem:Concavity}, applied to $h = \log F$ and $h = \log(1 - F)$.

\noindent
\textbf{Proof of (ii) $\Rightarrow$ (iii).} \
Suppose that $F$ is continuous on $\R$, differentiable on $J(F)$ with derivative $f = F'$ and satisfies the inequalities \eqref{eq:Characterize}. This implies that $h := f/F$ is non-increasing and $\tilde{h} := f/(1 - F)$ is non-decreasing on $J(F)$. For if $x, y \in J(F)$ with $x < y$, then by \eqref{eq:Characterize},
\begin{align*}
	\log F(x) \
	&\le \ \log F(y) + h(y) (x - y) \\
	&\le \ \log F(x) + h(x) (y - x) + h(y) (x - y) \\
	&= \ \log F(x) + \bigl( h(x) - h(y) \bigr) (y - x)
\end{align*}
and
\begin{align*}
	\log(1 - F(x)) \
	&\le \ \log(1 - F(y)) - \tilde{h}(y) (x - y) \\
	&\le \ \log(1 - F(x)) - \tilde{h}(x) (y - x) - \tilde{h}(y) (x - y) \\
	&= \ \log(1 - F(x)) + \bigl( \tilde{h}(y) - \tilde{h}(x) \bigr) (y - x) ,
\end{align*}
whence $h(x) \ge h(y)$ and $\tilde{h}(x) \le \tilde{h}(y)$.

\noindent
\textbf{Proof of (iii) $\Rightarrow$ (iv).} \
Suppose that $F$ satisfies the conditions in part~(iii). First of all this implies $f > 0$ on $J(F)$. Suppose $f(x_o) = 0$ for some $x_o \in J(F)$. Then isotonicity of $\tilde{h} = f/(1-F)$ implies $f(x) = 0$ for $x \le x_o$, and antitonicity of $h = f/F$ implies $f(x) = 0$ for $x \ge x_o$. Hence $F$ would be constant on $J(F)$, which violates that $F$ is a continuous distribution function on $\R$. Another consequence of these monotonicity properties is boundedness of $f$ on $J(F)$: If we fix any $x_o \in J(F)$, then for any other point $x \in J(F)$,
\[
	f(x) \ = \ \begin{cases}
		F(x) h(x) \ \le \ h(x_o)
		& \text{if} \ x \ge x_o , \\
		(1 - F(x)) \tilde{h}(x) \ \le \ \tilde{h}(x_o)
		& \text{if} \ x \le x_o .
	\end{cases}
\]
Finally, local Lipschitz-continuity of $f$ may be verified via Lemma~\ref{lem:Elementary}: Let $c, d \in J(F)$ with $c < d$. For arbitrary different $x,y \in (c,d)$,
\begin{align*}
	\frac{f(y) - f(x)}{y - x} \
	&=   \ \frac{F(y) h(y) - F(x) h(x)}{y - x} \\
	&=   \ h(y) \, \frac{F(y) - F(x)}{y - x} + F(x) \, \frac{h(y) - h(x)}{y - x} \\
	&\le \ h(c) \, \frac{F(y) - F(x)}{y - x} \\
	&\le \ h(c) \, \frac{\exp(h(x)(y - x)) - 1}{y - x} \, F(x) \\
	&\to \ h(c) h(x) F(x) \ \le \ h(c)^2 F(d)
\end{align*}
as $y \to x$. Hence
\begin{equation}
\label{eq:Lipschitz 1a}
	\limsup_{y \to x} \frac{f(y) - f(x)}{y - x}
	\ \le \ h(c)^2 F(d)
	\quad\text{for all} \ x \in (c,d)
\end{equation}
Analogously one can show that
\begin{equation}
\label{eq:Lipschitz 1b}
	\liminf_{y \to x} \frac{f(y) - f(x)}{y - x}
	\ \ge \ - \tilde{h}(d)^2 (1 - F(c))
	\quad\text{for all} \ x \in (c,d) .
\end{equation}
In particular, $f$ is Lipschitz-continuous on $(c,d)$ with Lipschitz-constant
\[
	\max \bigl\{ h(c)^2 F(d), \tilde{h}(d)^2 (1 - F(c)) \bigr\} .
\]
This proves local Lipschitz-continuity of $f$ on $J(F)$. In particular, $f$ is absolutely continuous with $L^1$-derivative $f'$. This means, $f'$ is a locally integrable function on $J(F)$ such that
\[
	f(y) - f(x) \ = \ \int_x^y f'(t) \, dt
	\quad\text{for all} \ x,y \in J(F) ,
\]
and it may be chosen such that
\[
	f'(x) \ \in \ \Bigl[ \liminf_{y \to x} \frac{f(y) - f(x)}{y - x},
		\limsup_{y \to x} \frac{f(y) - f(x)}{y - x} \Bigr]
\]
for any $x \in J(F)$. But for $c,d \in J(F)$ with $c < x < d$, the latter interval is contained in
\[
	\bigl[ - \tilde{h}(d)^2 (1 - F(c)), h(c)^2 F(d) \bigr]
	\ = \
	\Bigl[ \frac{- f(d)^2 (1 - F(c))}{(1 - F(d))^2}, \frac{f(c)^2 F(d)}{F(c)^2} \Bigr]
\]
according to \eqref{eq:Lipschitz 1a} and \eqref{eq:Lipschitz 1b}. Since $F$ and $f$ are continuous, letting $c,d \to x$ implies \eqref{eq:Second derivative}.

\noindent
\textbf{Proof of (iv) $\Rightarrow$ (i).} \
One can easily verify that a continuous distribution function $F$ is bi-log-concave if, and only if, $\log F$ and $\log(1 - F)$ are concave on $J(F)$. Hence (i) is a consequence of (iii), and it suffices to show that (iv) implies (iii).

According to Lemma~\ref{lem:Elementary}, $h$ is non-increasing on $J(F)$ if, and only if,
\[
	\limsup_{y \to x} \frac{h(y) - h(x)}{y - x} \ \le \ 0
\]
for any $x \in J(F)$. To verify this, let $y \in J(F) \setminus \{x\}$ and set $r := \min(x,y)$, $s := \max(x,y)$. Then it follows from \eqref{eq:Second derivative} and from continuity of $f$ that
\begin{align*}
	\frac{h(y) - h(x)}{y - x} \
	&=   \ \frac{f(y)/F(y) - f(x)/F(x)}{y - x} \\
	&=   \ \frac{1}{F(y)} \, \frac{f(y) - f(x)}{y - x}
		- \frac{f(x)}{F(x) F(y)} \, \frac{F(y) - F(x)}{y - s} \\
	&=   \ \frac{1}{F(y) (s - r)} \int_r^s f'(t) \, dt
		- \frac{f(x)}{F(x) F(y) (s - r)} \int_r^s f(t) \, dt \\
	&\le \ \frac{1}{F(y) (s - r)} \, \int_r^s \frac{f(t)^2}{F(t)} \, dt
		- \frac{f(x)}{F(x) F(y) (s - r)} \int_r^s f(t) \, dt \\
	&\to \ \frac{f(x)^2}{F(x)^2} - \frac{f(x)^2}{F(x)^2} \ = \ 0
\end{align*}
as $y \to x$.

Analogously one can show that $\tilde{h}$ is non-decreasing on $J(F)$.
\end{proof}

\begin{proof}[\bf Proof of \eqref{eq:MGF}]
For any fixed $x_o \in J(F)$, monotonicity of $f/F = \log(F)'$ implies that for $x \in J(F)$, $x < x_o$,
\[
	\frac{f}{F}(x) \ \ge \ \frac{\log F(x_o) - \log F(x)}{x - x_o} .
\]
Since $\log F(x) \to -\infty$ as $x \to \inf(J(F))$, this inequality implies that
\[
	T_1(F) = \sup_{x \in J(F)} \frac{f}{F}(x)
	= \lim_{x \to \inf(J(F))} \frac{f}{F}(x) \
	\begin{cases}
		> 0 , \\
		= \infty & \text{if} \ \inf(J(F)) > -\infty .
	\end{cases}
\]
Analogously one can show that
\[
	T_2(F) = \sup_{x \in J(F)} \frac{f}{1 - F}(x)
	= \lim_{x \to \sup(J(F))} \frac{f}{1 - F}(x) \
	\begin{cases}
		> 0 , \\
		= \infty & \text{if} \ \sup(J(F)) < \infty .
	\end{cases}
\]

For symmetry reasons it suffices to show that $\int e^{tx} F(dx)$ is finite for $t \in (0,T_2(F))$ and infinite for $t \ge T_2(F)$. Notice that for $t > 0$, Fubini's theorem yields
\begin{align*}
	\int e_{}^{tx} \, F(dx) \
	&= \ \int \int 1_{[z \le x]}^{} t e_{}^{tz} \, dz \, F(dx) \\
	&= \ t \int e_{}^{tz} (1 - F(z)) \, dz \\
	&= \ t \int \exp \bigl( tz + \log(1 - F(z)) \bigr) \, dz .
\end{align*}
In the case of $m := \sup(J(F)) < \infty$, the previous integral is smaller than $e^{tm} < \infty$ for $t < \infty = T_2(F)$. In the case of $m = \infty$, notice that $tz + \log(1 - F(z))$ is concave in $z \in \R$ with limit $- \infty$ as $z \to - \infty$. Thus the integral $\int e^{tx} \, F(dx)$ is finite if, and only if,
\[
	\lim_{z \to \infty} \, \frac{d}{dz} \bigl( tz + \log(1 - F(z)) \bigr)
	\ = \ \lim_{z \to \infty} \Bigl( t - \frac{f(z)}{1 - F(z)} \Bigr)
	\ = \ t - T_2(F)
\]
is strictly negative, which is equivalent to $t < T_2(F)$.
\end{proof}

\begin{proof}[\bf Proof of Lemma~\ref{lem:continuity.exponential.tails}]
The assertions are trivial if $L_n^o \equiv 1$ and $U_n^o \equiv 0$, meaning that no $G \in \FFblc$ fits in between $L_n$ and $U_n$. Otherwise let $G \in \FFblc$ such that $L_n \le G \le U_n$.

For part~(i) it suffices to show that for any $x \in J(G)$ the density $g = G'$ satisfies the inequality $g(x) \le \max\{\gamma_1,\gamma_2\}$. This is equivalent to Lipschitz-continuity of $G$ with the latter constant, and this property carries over to the pointwise infimum $L_n^o$ and supremum $U_n^o$. For $x \ge b$ it follows from concavity of $\log G$ and $G(a) \ge r$, $G(b) \le s$ that
\[
	g(x)
	\ \le \ \frac{g}{G}(x)
	\ \le \ \frac{g}{G}(b)
	\ \le \ \frac{\log G(b) - \log G(a)}{b - a}
	\ \le \ \frac{\log s - \log r}{b - a}
	\ = \ \gamma_1 .
\]
Similarly convexity of $- \log(1 - G)$ and the inequalities $G(a) \ge r$, $G(b) \le s$ imply that for $x \le a$,
\[
	g(x)
	\ \le \ \frac{g}{1 - G}(x)
	\ \le \ \frac{g}{1 - G}(a)
	\ \le \ \frac{- \log(1 - G(b)) + \log(1 - G(a))}{b - a}
	\ \le \ \gamma_2 .
\]
For $a < x < b$ we get the two inequalities
\[
	g(x)
	\ = \ G(x) \frac{g}{G}(x)
	\ \le \ G(x) \frac{\log G(x) - \log r}{x - a}
\]
and
\[
	g(x)
	\ = \ (1 - G(x)) \frac{g}{1-G}(x)
	\ \le \ (1 - G(x)) \frac{\log(1 - G(x)) - \log(1 - s)}{b - x} .
\]
The former inequality times $x - a$ plus the latter inequality times $b - x$ yields that
\[
	g(x)
	\ \le \ \frac{G(x) \log(G(x)/r)
		+ (1 - G(x)) \log \bigl( (1 - G(x))/(1 - s) \bigr)}{b - a}.
\]
But $h(y) := y \log(y/r) + (1 - y) \log ((1 - y)/(1 - s))$ is easily shown to be convex in $y \in (0,1)$, so
\[
	g(x) \ \le \ \max_{y = r, s} h(y) \ = \ \max\{\gamma_1,\gamma_2\}.
\]

As to part~(ii), it suffices to show that $G(x) \le G(a) \exp(\gamma_1(x - a))$ for $x \le a$ and $G(x) \ge 1 - (1 - G(b)) \exp(- \gamma_2(x - b))$ for $x \ge b$. We know from Theorem~\ref{thm:Characterize}~(ii) that this is true with $(g/G)(a)$ and $(g/(1 - G))(b)$ in place of $\gamma_1$ and $\gamma_2$, respectively. But it follows from $G(a) \le r$, $G(b) \ge s$ and concavity of $\log G$ that
\[
	\frac{g}{G}(a)
	\ \ge \ \frac{\log G(b) - \log G(a)}{b - a}
	\ \ge \ \frac{\log s - \log r}{b - a}
	\ = \ \gamma_1 ,
\]
while convexity of $- \log(1 - G)$ yields that $(g/(1 - G))(b) \ge \gamma_2$.
\end{proof}

\begin{proof}[\bf Proof of Theorem~\ref{thm:consistency}]
Suppose that $F \not\in \FFblc$, that means, $\log F$ or $\log(1 - F)$ is not concave. In the former case there exist real numbers $x_0 < x_1 < x_2$ such that $\log F(x_1) < (1 - \lambda) \log F(x_0) + \lambda \log F(x_2)$, where $\lambda := (x_1 - x_0)/(x_2 - x_0) \in (0,1)$. Then with probability tending to one, $\log U_n(x_1) < (1 - \lambda) \log L_n(x_0) + \lambda \log L_n(x_2)$, whence no log-concave distribution function fits between $L_n$ and $U_n$. Analogous arguments apply in the case of $\log(1 - F)$ violating concavity.

Now suppose that $F \in \FFblc$. Obviously, $\Pr(L_n^o \le U_n^o) \ge \Pr(L_n \le F \le U_n) \ge 1 - \alpha$. Since $L_n$ and $U_n$ are assumed to be non-decreasing, and since $F$ is continuous, a standard argument shows that pointwise convergence implies uniform convergence in probability, i.e.\ $\|L_n - F\|_\infty \to_p 0$ and $\|U_n - F\|_\infty \to_p 0$. This implies that
\begin{equation}
\label{eq:consistencyG}
	\sup_{G \in \FFblc \,:\, L_n \le G \le U_n} \, \|G - F\|_\infty
	\ \le \ \|L_n - F\|_\infty + \|U_n - F\|_\infty
	\ \to_p \ 0 ,
\end{equation}
because $L_n \le L_n^o \le U_n^o \le U_n$ in the case of $L_n^o \le U_n^o$.

Now let $K$ be a compact subset of $J(F)$, and let $h_G := \log(G)'$ for $G \in \FFblc$. Since $h_F = f/F$ is continuous and non-increasing on $J(F)$, for any fixed $\eps > 0$ there exist points $a_0 < a_1 < \cdots < a_m < a_{m+1}$ in $J(F)$ such that $K \subset [a_1,a_m]$ and
\[
	0 \ \le \ h_F(a_{i-1}) - h_F(a_i) \ \le \ \eps
	\quad\text{for} \ 1 \le i \le m+1 .
\]
For $G \in \FFblc$ with $L_n \le G \le U_n$, for any $x \in K$ it follows from monotonicity of $h_F$ and $h_G$ that
\begin{align*}
	\sup_{x \in K} \bigl( h_G(x) - h_F(x) \bigr) \
	&\le \ \max_{i=1,\ldots,m-1} \bigl( h_G(a_i) - h_F(a_{i+1}) \bigr) \\
	&\le \ \max_{i=1,\ldots,m-1} \Bigl(
		\frac{\log G(a_i) - \log G(a_{i-1})}{a_i - a_{i-1}} - h_F(a_{i+1})
			\Bigr) \\
	&\le \ \max_{i=1,\ldots,m-1} \Bigl(
		\frac{\log U_n(a_i) - \log L_n(a_{i-1})}{a_i - a_{i-1}} - h_F(a_{i+1})
			\Bigr) \\
	&= \ \max_{i=1,\ldots,m-1} \Bigl(
		\frac{\log F(a_i) - \log F(a_{i-1})}{a_i - a_{i-1}} - h_F(a_{i+1})
			\Bigr) + o_p(1) \\
	&\le \ \max_{i=1,\ldots,m-1} \bigl( h_F(a_{i-1}) - h_F(a_{i+1}) \bigr)
		+ o_p(1) \\
	&\le \ 2\eps + o_p(1) .
\end{align*}
Analogously,
\begin{align*}
	\sup_{x \in K} \bigl( h_F(x) - h_G(x) \bigr) \
	&\le \ \max_{i=1,\ldots,m-1} \bigl( h_F(a_i) - h_F(a_{i+2}) \bigr)
		+ o_p(1) \\
	&\le \ 2\eps + o_p(1) .	
\end{align*}
Since $\eps > 0$ is arbitrarily small, this shows that
\begin{equation}
\label{eq:consistencyG'}
	\sup_{G \in \FFblc \,:\, L_n \le G \le U_N}
		\bigl\| \log(G)' - \log(F)' \bigr\|_{K,\infty} \ = \ o_p(1) .
\end{equation}
Analogously one can show that
\[
	\sup_{G \in \FFblc \,:\, L_n \le G \le U_N}
		\bigl\| \log(1 - G)' - \log(1 - F)' \bigr\|_{K,\infty} \ = \ o_p(1) .
\]
Moreover, since $G' = \log(G)' \, G$, it follows from \eqref{eq:consistencyG} and \eqref{eq:consistencyG'} that
\[
	\sup_{G \in \FFblc \,:\, L_n \le G \le U_N}
		\|G' - F'\|_{K,\infty} \ = \ o_p(1) .
\]

Finally, let $x_1 < \sup(J(F))$ and $b_1 < f(x_1)/F(x_1)$. As in the proof of Lemma~\ref{lem:continuity.exponential.tails}~(ii) one may argue that for any fixed $x_1' > x_1$, $x_1' \in J(F)$,
\[
	U_n^o(x) \ \le \ U_n(x')
		\exp \Bigl( \frac{\log L_n(x_1') - \log U_n(x_1)}{x_1' - x_1}
			(x - x') \Bigr)
\]
for all $x \le x' \le x_1$. But
\[
	\frac{\log L_n(x_1') - \log U_n(x_1)}{x_1' - x_1}
	\ \to_p \ \frac{\log F(x_1') - \log F(x_1)}{x_1' - x_1}
	\ > \ b_1
\]
if $x_1 \le \inf(J(F))$ or $x_1'$ is sufficiently close to $x_1 \in J(F)$. This shows that with asymptotic probability one,
\[
	U_n^o(x) \ \le \ U_n(x') \exp( b_1 (x - x'))
\]
for all $x \le x' \le x_1$. Analogously one can prove the claim about $1 - L_n^o$ on halflines $[x_2,\infty)$, $x_2 > \inf(J(F))$.
\end{proof}

\begin{proof}[\bf Proof of Corollary~\ref{cor:consistency}]
Without loss of generality let $0 \in J(F)$; otherwise we could shift the coordinate system suitably and adjust the constant $a$ in our bound for $|\phi'|$. Notice that for any $z \in \R$,
\[
	\phi(z) - \phi(0)
	\ = \ \int_{-\infty}^\infty
		\bigl( 1_{[0 \le x < z]} - 1_{[z \le x < 0]} \bigr) \phi'(x) \, dx ,
\]
so by Fubini's theorem,
\[
	\int \phi \, dG
	\ = \ \phi(0)
		+ \int_{\R} \phi'(x) \bigl( 1_{[x \ge 0]} - G(x) \bigr) \, dx ,
\]
provided that
\begin{equation}
\label{ineq:G.phi}
	\int |\phi'(x)| \bigl| 1_{[x \ge 0]} - G(x) \bigr| \, dx \ < \ \infty .
\end{equation}

By assumption, for arbitrary numbers $b_1' \in (0,T_1(F))$ and $b_2' \in (0,T_2(F))$ there exist points $x_1, x_2 \in J(F)$ with $x_1 \le 0 \le x_2$ and
\[
	f(x_1)/F(x_1) \ > \ b_1', \quad
	f(x_2)/(1 - F(x_2)) \ > \ b_2' .
\]
Then it follows from Theorem~\ref{thm:consistency}~(ii) that with asymptotic probability one,
\begin{align}
\label{ineq:as.tail.U}
	U_n^o(x) \
	&\le \ U_n(x') \exp(b_1'(x - x'))
		\quad\text{for} \ x \le x' \le x_1
\intertext{and}
\label{ineq:as.tail.L}
	1 - L_n^o(x) \
	&\le \ (1 - L_n(x')) \exp(- b_2'(x - x'))
		\quad\text{for} \ x \ge x' \ge x_2 .
\end{align}
If we choose $b_1' > b_1$ and $b_2' > b_2$, the inequalities \eqref{ineq:as.tail.U} and \eqref{ineq:as.tail.L} imply \eqref{ineq:G.phi} for arbitrary distribution functions $G$ with $L_n^o \le G \le U_n^o$. More precisely, for any fixed $c \ge 0$ and $\delta := \min\{b_1' - b_1, b_2' - b_2\} > 0$,
\begin{align*}
	\int_{-\infty}^{x_1-c} |\phi'(x)| U_n^o(x) \, dx \
	&\le \ U_n(x_1)
		\int_{-\infty}^{x_1-c} \exp(a - b_1 x + b_1'(x - x_1)) \, dx \\
	&\le \ U_n(x_1) \exp(a - b_1 x_1 - \delta c)
		\int_{-\infty}^0 \exp(\delta y) \, dy \\
	&= \ \frac{U_n(x_1) \exp(a - b_1 x_1 - \delta c)}{\delta}
\end{align*}
and
\[
	\int_{x_2 + c}^\infty |\phi'(x)| (1 - L_n^o(x)) \, dx
	\ \le \ \frac{(1 - L_n(x_1)) \exp(a + b_2 x_2 - \delta c)}{\delta} .
\]
The same inequalities hold if $L_n$, $U_n$, $L_n^o$ and $U_n^o$ are all replaced with $F$. Thus
\begin{equation}
\label{eq:int.phi.dG}
	\sup_{G \,:\, L_n^o \le G \le U_n^o}
		\Bigl| \int \phi \, dG - \int \phi \, dF \Bigr|
	\ = \ \sup_{G \,:\, L_n^o \le G \le U_n^o}
		\Bigl| \int_{-\infty}^\infty \phi'(x) (F - G)(x) \, dx \Bigr| \\
\end{equation}
is not larger than
\begin{align*}
	\sup_{G \,:\, L_n^o \le G \le U_n^o}
	& \|G - F\|_\infty
		\int_{x_1-c}^{x_2+c} |\phi'(x)| \, dx \\
	&+ \ \int_{-\infty}^{x_1-c} |\phi'(x)| (U_n^o + F)(x) \, dx
		+ \int_{x_2+c}^\infty |\phi'(x)| (2 - L_n^o - F)(x) \, dx \\
	\le \ & \frac{2 F(x_1) \exp(a - b_1 x_1 - \delta c)}{\delta}
		+ \frac{2 (1 - F(x_2)) \exp(a + b_2 x_2 - \delta c)}{\delta} + o_p(1) .
\end{align*}
But the limit on the right hand side becomes arbitrarily small for sufficiently large $c > 0$.
\end{proof}

\begin{proof}[\bf Proof of Theorem~\ref{thm:consistency.rates}]
It follows from standard results about the empirical process on the real line that for any fixed $\eps \in (0,1)$ there exists a constant $\kappa_\eps > 0$ such that with probability at least $1 - \eps$,
\[
	|\hat{F}_n - F| \ \le \ \kappa_\eps n^{-1/2} (F (1 - F))^\gamma
\]
on $\R$. Let us assume that the previous inequalities hold and that $L_n^o \le U_n^o$.

For a constant $\lambda_\eps > 0$ to be specified later it follows from $\lambda_\eps n^{-1/(2 - 2\gamma)} \le F \le 1 - \lambda_\eps n^{-1/(2 - 2\gamma)}$ that
\[
	\hat{F}_n
	\ \ge \ \Bigl( 1 - \frac{|\hat{F}_n - F|}{F} \Bigr) F
	\ \ge \ (1 - \kappa_\eps \lambda_\eps^{\gamma - 1}) \lambda_\eps
		n^{-1/(2 - 2\gamma)}
	\ = \ (\lambda_\eps - \kappa_\eps \lambda_\eps^\gamma) n^{-1/(2 - 2\gamma)}
\]
and
\[
	1 - \hat{F}_n
	\ \ge \ (\lambda_\eps - \mu_\eps \lambda_\eps^\gamma) n^{-1/(2 - 2\gamma)} .
\]
Thus we choose $\lambda_\eps$ sufficiently large such that the number $\lambda_\eps - \kappa_\eps \lambda_\eps^\gamma$ exceeds $\lambda$. Then the interval
\[
	J_n
	\ := \ \{\lambda_\eps n^{-1/(2 - 2\gamma)}
		\le F \le 1 - \lambda_\eps n^{-1/(2 - 2\gamma)}\}
\]
is a subset of $\{\lambda n^{-1/(2 - 2\gamma)} \le \hat{F}_n \le 1 - \lambda n^{-1/(2 - 2\gamma)}\}$. On this interval $J_n$,
\[
	\frac{\hat{F}_n (1 - \hat{F}_n)}{F(1-F)}
	\ \le \ \max \Bigl\{ \frac{\hat{F}_n}{F}, \frac{1 - \hat{F}_n}{1-F} \Bigr\}
	\ \le \ 1 + \frac{|\hat{F}_n - F|}{\min(F,1-F)}
	\ \le \ 1 + \kappa_\eps \lambda_\eps^{\gamma-1} ,
\]
and for any function $h$ with $L_n \le h \le U_n$,
\begin{equation}
\label{ineq:central.part}
	\frac{|h - F|}{(F(1-F))^\gamma}
	\ \le \ \frac{|h - \hat{F}_n|}{(\hat{F}_n (1 - \hat{F}_n))^\gamma}
		\Bigl( \frac{\hat{F}_n (1 - \hat{F}_n)}{F(1-F)} \Bigr)^\gamma
		+ \frac{|\hat{F}_n - F|}{(F(1-F))^\gamma}
	\ \le \ \nu_\eps n^{-1/2}
\end{equation}
with $\nu_\eps := \kappa (1 + \kappa_\eps \lambda_\eps^{\gamma - 1})^\gamma + \kappa_\eps$. In particular, the boundaries $L_n$ and $U_n$ themselves satisfy \eqref{ineq:central.part} on $J_n$.

Again we assume without loss of generality that $0 \in J(F)$. For arbitrary fixed numbers $b_1' \in (0,T_1(F))$ and $b_2' \in (0,T_2(F))$ we choose points $x_1, x_2 \in J(F)$ with $x_1 < 0 < x_2$ such that $f(x_1)/F(x_1) > b_1'$ and $f(x_2)/(1 - F(x_2)) > b_2'$. For sufficiently large $n$, $[x_1, x_2] \subset J_n$, and we may even assume that \eqref{ineq:as.tail.U} and \eqref{ineq:as.tail.L} are satisfied, too. Writing $J_n = [x_{n1},x_{n2}]$, we may deduce from \eqref{eq:int.phi.dG} and \eqref{ineq:central.part} that
\begin{align*}
	\sup_{G \,:\, L_n^o \le G \le U_n^o} \Bigl| \int \phi \, d(G - F) \Bigr| \
	\le \ &\nu_\eps n^{-1/2}
		\int_{x_{n1}}^{x_{n2}} |\phi'(x)| F(x)^\gamma (1 - F(x))^\gamma \, dx \\
		&+ \ \int_{-\infty}^{x_{n1}} |\phi'(x)| (F + U_n^o)(x) \, dx \\
		&+ \ \int_{x_{n2}}^\infty |\phi'(x)| (2 - F - L_n^o)(x) \, dx .
\end{align*}
Notice that
\begin{align*}
	F(x) \
	&\le \ F(x_1) \exp(b_1'(x - x_1))
		\quad\text{for} \ x \le x_1 , \\
	1 - F(x) \
	&\le \ (1 - F(x_2)) \exp(- b_2'(x - x_2))
		\quad\text{for} \ x \ge x_2 .
\end{align*}
In particular, for $x = x_{n1}, x_{n2}$ it follows from these inequalities and $F(x_{n1}) = 1 - F(x_{n2}) = \lambda_\eps n^{-1/(2 - 2\gamma)}$ that
\begin{equation}
\label{ineq:xn}
	x_{n1} \ \ge \ O(1) - \frac{\log n}{b_1'(2 - 2\gamma)}
	\quad\text{and}\quad
	x_{n2} \ \le \ O(1) + \frac{\log n}{b_2'(2 - 2\gamma)} .
\end{equation}
Notice also that by \eqref{ineq:as.tail.L}, \eqref{ineq:as.tail.U} and \eqref{ineq:central.part},
\begin{align*}
	(F + U_n^o)(x) \
	&\le \ (F + U_n^o)(x_{n1}) \exp(b_1'(x - x_{n1})) \\
	&\le \ \omega_\eps n^{-1/(2 - 2\gamma)} \exp(b_1'(x - x_{n1}))
		\quad\text{for} \ x \le x_{n1} , \\
	(2 - F - L_n^o)(x) \
	&\le \ \omega_\eps n^{-1/(2 - 2\gamma)} \exp(- b_2'(x - x_{n2}))
		\quad\text{for} \ x \ge x_{n2} ,
\end{align*}
where $\omega_\eps := \lambda_\eps + \nu_\eps \lambda_\eps^\gamma$. These considerations show that
\[
	\sup_{G \,:\, L_n^o \le G \le U_n^o} \Bigl| \int \phi \, d(G - F) \Bigr|
	\ \le \ I_{n0} + I_{n1} + I_{n1}' + I_{n2} + I_{n2}'
\]
with
\begin{align*}
	I_{n0} \
	&:= \ \nu_\eps n^{-1/2}
			\int_{x_1}^{x_2} |\phi'(x)| \, dx
		\ = \ O(n^{-1/2}) , \\
	I_{n1} \
	&:= \ \nu_\eps n^{-1/2}
			\int_{x_{n1}}^{x_1} |\phi'(x)| F(x)^\gamma \, dx
		\ = \ O \biggl( n^{-1/2}
			\int_{x_{n1}}^{x_1} |\phi'(x)| e^{\gamma b_1'x} \, dx
			\biggr) , \\
	I_{n1}' \
	&:= \ \int_{-\infty}^{x_{n1}} |\phi'(x)| (F + U_n^o)(x) \, dx
		\ = \ O \biggl( n^{-1/(2 - 2\gamma)}
			\int_{-\infty}^{x_{n1}} |\phi'(x)| e^{b_1'(x - x_{n1})} \, dx
			\biggr) , \\
	I_{n2} \
	&:= \ \nu_\eps n^{-1/2}
			\int_{x_2}^{x_{n2}} |\phi'(x)| (1 - F(x))^\gamma \, dx
		\ = \ O \biggl( n^{-1/2}
			\int_{x_2}^{x_{n2}} |\phi'(x)| e^{- \gamma b_2'x} \, dx
			\biggr) , \\
	I_{n2}' \
	&:= \ \int_{x_{n2}}^\infty |\phi'(x)| (2 - F - L_n^o)(x) \, dx
		\ = \ O \biggl( n^{-1/(2 - 2\gamma)}
			\int_{x_{n2}}^\infty |\phi'(x)| e^{-b_2'(x - x_{n2})} \, dx
			\biggr) .
\end{align*}

As to part~(i), suppose that $|\phi'(x)| \le a (1 + |x|^{k-1})$ for arbitrary $x \in \R$ and some constant $a > 0$. Then both $I_{n1}$ and $I_{n2}$ are of order
\begin{align*}
	O \biggl( n^{-1/2} \int_0^{O(\log n)}
		(1 + s^{k-1}) \exp(- \gamma b' s) \, ds \biggr)
	\ = \ \begin{cases}
		O(n^{-1/2}) & \text{if} \ \gamma > 0 , \\
		O(n^{-1/2} (\log n)^k) & \text{if} \ \gamma = 0 ,
	\end{cases}
\end{align*}
where $b' := \min\{b_1',b_2'\} > 0$. Moreover, both $I_{n1}'$ and $I_{n2}'$ are of order
\begin{align*}
	O \biggl( n^{-1/(2 - 2\gamma)}
		\int_0^\infty O \bigl( (\log n)^{k-1} + s^{k-1} \bigr) e^{-b's} \, ds
		\biggr) \
	&= \ O \bigl( n^{-1/(2 - 2\gamma)} (\log n)^{k-1} \bigr) \\
	&= \ \begin{cases}
		o(n^{-1/2}) & \text{if} \ \gamma > 0 , \\
		O(n^{-1/2} (\log n)^{k-1}) & \text{if} \ \gamma = 0 .
	\end{cases}
\end{align*}
This proves the assertion in part~(i).

For functions $\phi$ as in part~(ii), let $b_1' > b_1$ and $b_2' > b_2$ such that $b_1 \ne \gamma b_1'$ and $b_2 \ne \gamma b_2'$. Then
\[
	I_{n1}
	\ = \ O \biggl( n^{-1/2}
		\int_0^{O(1) + (\log n)/(b_1'(2 - 2\gamma))}
			\exp((b_1 - \gamma b_1') s) \, ds \biggr)
	\ = \ O(n^{-\beta_1})
\]
with
\[
	\beta_1 \ := \ \frac{1}{2} - \frac{(b_1 - \gamma b_1')^+}{b_1' (2 - 2\gamma)}
	\ = \ \frac{1 - \gamma - (b_1/b_1' - \gamma)^+}{2(1 - \gamma)}
	\ = \ \frac{1 - \max(b_1/b_1',\gamma)}{2(1 - \gamma)} ,
\]
and
\[
	I_{n2} \ = \ O(n^{-\beta_2})
	\quad\text{with}\quad
	\beta_2 \ := \ \frac{1 - \max(b_2/b_2',\gamma)}{2(1 - \gamma)} .
\]
Furthermore,
\begin{align*}
	I_{n1}' \
	&= \ O \biggl( n^{-1/(2-2\gamma)}
			\int_{-\infty}^{x_{n1}} \exp(- b_1x + b_1'(x - x_{n1})) \, dx \biggr) \\
	&= \ O \biggl( n^{-1/(2-2\gamma)} \exp(- b_1 x_{n1})
			\int_0^\infty \exp(- (b_1' - b_1) s) \, ds \biggr) \\
	&= \ O \bigl( n^{-1/(2 - 2\gamma)} \exp(- b_1 x_{n1}) \bigr)
		\ = \ O \bigl( n^{-(1 - b_1/b_1')/(2 - 2\gamma)} \bigr)
		\ = \ O(n^{-\beta_1})
\end{align*}
and
\[
	I_{n2}' = \ O(n^{-\beta_2}) .
\]
This proves the assertion in part~(ii). If $\tilde{\gamma} := \max \bigl\{ b_1/T_1(F), b_2/T_2(F) \bigr\} < \gamma$, we may choose $b_1'$ and $b_2'$ such that $b_1/b_1', b_2/b_2' < \gamma$, resulting in $\beta_1 = \beta_2 = 1/2$. If $\tilde{\gamma} \ge \gamma$, the exponents $\beta_1, \beta_2$ are strictly smaller than but arbitrarily close to $(1 - \tilde{\gamma})/(2(1 - \gamma))$.
\end{proof}

\paragraph{Acknowledgement.}
Constructive comments by two referees are gratefully acknowledged.



\begin{thebibliography}{77}
\bibitem{Bagnoli_Bergstrom_2005}
	\textsc{Bagnoli, M.} and \textsc{Bergstrom, T.} (2005).
	\newblock Log-concave probability and its applications.
	\newblock \textsl{Econ.\ Theory \textbf{26}}, 445-469.
\bibitem{Bahadur_Savage_1956}
	\textsc{Bahadur, R.R.} and \textsc{Savage, L.J.} (1956).
	\newblock The nonexistence of certain statistical procedures
		in nonparametric problems.
	\newblock \textsl{Ann.\ Math.\ Statist.\ \textbf{27}}, 1115-1122.
\bibitem{Berk_Jones_1979}
	\textsc{Berk, R.H.} and \textsc{Jones, D.H.} (1979).
	\newblock Goodness-of-fit statistics that dominate
		the Kolmogorov statistics.
	\newblock \textsl{Z.\ Wahrsch.\ Verw.\ Gebiete \textbf{47}}, 47-59.
\bibitem{Cule_etal_2010}
	\textsc{Cule, M.L.}, \textsc{Samworth, R.J.} and
		\textsc{Stewart, M.I.} (2010).
	\newblock Maximum likelihood estimation of a multidimensional
		log-concave density (with discussion).
	\newblock \textsl{J.\ Roy.\ Statist.\ Soc., Ser.\ B \textbf{72}}, 545-600.
\bibitem{Duembgen_1998}
	\textsc{D\"umbgen, L.} (1998).
	\newblock New goodness-of-fit tests and their application to
		nonparametric confidence sets.
	\newblock \textsl{Ann.\ Statist.\ \textbf{26}(1)}, 288-314.
\bibitem{Duembgen_Rufibach_2009}
	\textsc{D\"umbgen, L.} and \textsc{Rufibach, K.} (2009).
	\newblock Maximum likelihood estimation of a log-concave density and
		its distribution function: basic properties and uniform consistency.
	\newblock \textsl{Bernoulli \textbf{15}(1)}, 40-68.
\bibitem{Duembgen_Rufibach_2011}
	\textsc{D\"umbgen, L.} and \textsc{Rufibach, K.} (2011).
	\newblock logcondens: Computations related to univariate
		log-concave density estimation.
	\newblock \textsl{J.\ Statist.\ Software \textbf{39}(6)}.
\bibitem{Duembgen_etal_2011}
	\textsc{D\"umbgen, L.}, \textsc{Samworth, R.J.} and
		\textsc{Schuhmacher, D.} (2011).
	\newblock Approximation by log-concave distributions,
		with applications to regression.
	\newblock \textsl{Ann.\ Statist.\ \textbf{39}(2)}, 702-730.
\bibitem{Duembgen_Wellner_2014}
	\textsc{D\"umbgen, L.} and \textsc{Wellner, J.A.} (2014).
	\newblock Confidence bands for distribution functions:
		A new look at the law of the iterated logarithm.
	\newblock Preprint.
\bibitem{Kleiber_Kotz_2003}
	\textsc{Kleiber, C.} and \textsc{Kotz, S.} (2003).
	\newblock \textsl{Statistical size distributions in
		economics and actuarial sciences.}
	\newblock Wiley Series in Probability and Statistics.
\bibitem{Massart_1990}
	\textsc{Massart, P.} (1990).
	\newblock The tight constant in the Dvoretzky-Kiefer-Wolfowitz Inequality.
	\newblock \textsl{Ann.\ Probab.\ \textbf{18}}, 1269-1283.
\bibitem{Owen_1995}
	\textsc{Owen, A.B.} (1995).
	\newblock Nonparametric likelihood confidence bands for a
		distribution function.
	\newblock \textsl{J.\ Amer.\ Statist.\ Assoc.\ \textbf{90}(430)}, 516-521.
\bibitem{Robertson_etal_1988}
	\textsc{Robertson, T.}, \textsc{Wright, F.T.} and \textsc{Dykstra, R.L.} (1988).
	\newblock \textsl{Order restricted statistical inference},
	\newblock Wiley \& Sons.
\bibitem{Seregin_Wellner_2010}
	\textsc{Seregin, A.} and \textsc{Wellner, J.A.} (2010).
	\newblock Nonparametric estimation of multivariate
		convex-transformed densities.
	\newblock \textsl{Ann.\ Statist.\ \textbf{38}(6)}, 3751-3781.
\bibitem{Schuhmacher_etal_2011}
	\textsc{Schuhmacher, D.}, \textsc{H\"usler, A.} and
		\textsc{D\"umbgen, L.} (2011).
	\newblock Log-concave distributions as a nearly parametric model.
	\newblock \textsl{Statist.\ Risk Model.\ \textbf{28}(3)}, 277-295.
\bibitem{Shorack_Wellner_1986}
	\textsc{Shorack, G.R.} and \textsc{Wellner, J.A.} (1986).
	\newblock \textsl{Empirical processes with applications to statistics.}
	\newblock Wiley, New York.
\bibitem{Walther_2009}
	\textsc{Walther, G.} (2009).
	\newblock Inference and modeling with log-concave distributions.
	\newblock \textsl{Statist.\ Sci.\ \textbf{24}}, 319-327.
\bibitem{Wellner_Jager_2007}
	\textsc{Wellner, J.A.} and \textsc{Jager, L.} (2007).
	\newblock Goodness-of-fit tests via phi-divergences.
	\newblock \textsl{Ann.\ Statist.\ \textbf{35}(5)}, 2018-2053.
\bibitem{Woolridge_2000}
	\textsc{Woolridge, J.M.} (2000).
	\newblock CEOSAL2 (Instructional Stata data sets for Econometrics).
	\newblock Boston College Department of Economics.
\end{thebibliography}
\end{document}